\documentclass{article}
\usepackage[utf8]{inputenc}
\usepackage{graphicx}
\usepackage{authblk}
\usepackage{booktabs}
\usepackage{graphicx}
\usepackage{verbatim}
\usepackage{url}
\usepackage{caption}
\usepackage{subcaption}
\usepackage{booktabs}

\title{A Similarity Approach to Cities and Features}

\author[1]{Luciano da F. Costa}
\author[1]{Eric K. Tokuda  }

\affil[1]{
   São Carlos Institute of Physics - DFCM \protect\\
University of São Paulo \protect\\
P.O. Box 369, São Carlos, S.P. \protect\\
13560-970 Brazil
}

\date{2nd Feb. 2022}
\begin{document}

    \maketitle
    \begin{abstract}
Characterizing the structure of cities constitutes an important
task since the identification of similar cities can promote
sharing of respective experiences.  In the present work, we 
consider 20 European cities from 5 respective countries and with comparable populations, each
of which characterized in terms of four topological as
well as one geometrical feature.  These cities are then mapped
into respective networks by considering their
pairwise similarity as gauged by the coincidence methodology,
which consists of combining the Jaccard and interiority indices.
The methodology incorporates a parameter alpha that can control the relative contribution of features with the same or
opposite signs to the overall similarity.  Interestingly, the
maximum modularity cities network is obtained for a non-standard
parameter configuration, showing that it could not be
obtained were not for the adoption of the parameter alpha.  The network with maximum modularity presents four communities that can be directly related to four of the five considered countries, corroborating not only the effectiveness of the adopted features and similarity methodology, but also indicating a surprising tendency of the cities from a same country of being similar, while differing from cities from other countries.  The coincidence methodology was then applied in order to investigate the effect of several features combinations on the respectively obtained networks, leading to a highly modular features network containing four main communities that can be understood as the main possible models for the considered cities.
    \end{abstract}

\maketitle

\section{Introduction}

The physical world is intrinsically characterized by unending diversity.  Towns and cities are no exception.  As a consequence of numerous influences -- including geography, history, economy, age, climate, as well as traditions and culture -- it is completely impossible to find two towns or cities that are identical regarding all their respective geometrical and topological characteristics, or \emph{features}.

Given that cities are dynamic structures, accommodating continuously across time and space, it becomes a important task not only to characterize them according to a representative number of features, but also to devise and apply the means for quantifying, in pairwise fashion, how much cities are similar~\cite{costa2010efficiency,batty2013new,strano2013urban,domingues2018topological,viana2011fast,travenccolo2008hierarchical,barthelemy2016structure}.  These initiatives can lead to several valuable results.  For instance, cities that
are found to be similar can consider sharing their respective administrative and
planning experiences.  In addition, as soon as the pairwise similarity between cities has been effectively quantified, it becomes possible to generate respective networks in which each city becomes a node, while the interconnections reflect the respective similarities.  Analysis of these networks by using the wealthy of concepts and methods from network science (e.g.~\cite{boccaletti2006complex}) can then provide an ample understanding about types of cities corresponding to large network communities, as well as more specific types of cities.  

While city networks may present common features,  it is possible to group city networks into historical non-planned cities and modern planned cities~\cite{crucitti2006centrality}.  The authors of~\cite{rosvall2005networks} employ concepts from information science to compare cities. They map the city street network to an information network and they consider a complexity measurement $S$ for locating streets in this network. They observed that modern cities, such as Manhattan, tend to be more organized (easier to navigate) than older cities as Umeå.

In~\cite{crucitti2006centrality,strano2013urban}, the authors employ network centrality measurements for city comparison. They calculated four centrality measures, namely closeness, betweenness, straightness, and information, from network patches to compare cities from different countries. In~\cite{crucitti2006centrality}, the result indicates that self-organized cities tend to exhibit different centrality values when compared to more planned ones. In~\cite{strano2013urban}, besides the differences perceived by the accessibility measurements, they also observed heterogeneity in the street length distribution across cities.

While the traditional use of networks for studying cities often takes into account their street networks, it is also possible to model the economical relationships between cities (and its service firms) through a cities network~\cite{taylor2001specification}. In this model, differently from street networks, each node correspond to a city and the edges correspond to the relation between cities.
In~\cite{zhou2014recognizing}, the authors also consider cities as nodes of a network, but here the connectivity is based on the city identity recognition based on geolocated images. The authors argue that, in general, the city can be characterized by observers who look at a set of respective images. The obtained network quantifies the visual similarity among cities. A related approach was proposed by~\cite{dobesova2019similarity}, where a comparative analysis of cities based on the land use distribution was performed. The author collected data from 100 cities, computed image descriptors and compared the images using a hierarchical clustering algorithm. 
In~\cite{preoctiuc2013exploring}, the authors used venue-based data from social networks to compare cities.

Similarity indices (e.g.~\cite{vijaymeena2016survey,mirkin1996mathematical,akbas2014l1,costa2021onsimilarity}) have been frequently used in scientific and technological applications.
Given that the cosine similarity and Pearson correlation coefficient can cope with
real-valued features (e.g.~\cite{costa2021onsimilarity,costa2011analyzing}), these two similarity measurements correspond to frequently employed approaches for transforming datasets into respective graphs (or networks, e.g.~\cite{costa2021coincidence}).

Despite its good potential for similarity characterization, the Jaccard index (e.g.~\cite{costa2021onsimilarity,vijaymeena2016survey} defined in Eq.~\ref{eq:jaccard} respectively to two sets $A$ and $B$, has been largely restricted to the treatment of categorical or binary data.  
\begin{equation}
    \mathcal{J}(A,B) = \frac{|A \cap B|}{|A \cup B|}
    \label{eq:jaccard}
\end{equation}

where $|A| > 0$ stands for the \emph{cardinality}, or number of elements, in set $A$.
The Jaccard index has also found not to be able to take into account how much each of the compared sets is interior to the other~\cite{costa2021further,costa2021onsimilarity}.  

A generalization of the Jaccard index capable of taking into account real, possibly negative valued vectors or functions, as well as incorporating the quantification of the relative interiority between the two compared vectors has been described in~\cite{costa2021further,costa2021onsimilarity}.  More specifically, multiset principles~\cite{costa2021multiset} are used to translate the intersection and union involving real values into respective multiset operations involving the minimum and maximum functions, as well as functions indicating the sign of the operands~\cite{costa2021onsimilarity}.  In order to incorporate information about the interiority between the two compared vectors, the interiority (or overlap index~\cite{vijaymeena2016survey}) is also calculated by using multiset representation and then multiplied by the Jaccard index, resulting in the \emph{coincidence similarity index}~\cite{costa2021further,costa2021onsimilarity}.

In addition to being applicable to real values and incorporating information about the relative interiority of the vectors, the coincidence index has also been shown to be closely related to the generalized Kronecker delta function, capable of implementing the most strict comparison of similarity between two numeric values~\cite{costa2021onsimilarity}.

As a consequence of the aforementioned characteristics, the coincidence index was shown~\cite{costa2021onsimilarity}, while comparing vectors, to lead to results that are substantially more strict than the cosine similarity and Pearson correlation alternative approaches.  These properties have allowed the coincidence index to be applied as a mean to translate datasets described by respective features, into graphs or networks presenting a marked level of interconnectivity detail as well as enhanced modularity~\cite{costa2021onsimilarity,costa2021coincidence}.

Interestingly, the own coincidence method for translating datasets into
networks can be employed in order to approach another important and challenging problem in pattern recognition, namely the objective quantitative identification of the effect that different choices of features can have on the resulting networks, a problem directly related to feature analysis and selections (e.g.~\cite{costa2021elementary}).  This type of study is essential for supplying valuable information, not only how features are interrelated, but also to select particularly suitable features then used to obtain network representations.

Given a dataset and a set of respective features, the basic idea, as described in~\cite{costa2021elementary}, consists in deriving the \emph{features network} respectively implied by each possible (or of interest) feature combination, and then to consider the obtained weight matrices as features to obtain a network where each of those networks is associated to a node, while the pairwise interconnections reflect the respective similarity as gauged by the coincidence index.  
The present work aims at addressing the interesting problem of comparing several cities worldwide in terms of networks obtained by the coincidence methodology~\cite{costa2021coincidence}.  The primary motivation for this study consists in harnessing more detailed and modular description of cities relationships as allowed by the intrinsic ability of the coincidence similarity index in promoting more detailed and strict information about the compared entities.

Twenty European cities with similar populations have been arbitrarily chosen, corresponding to four cities from each of 5 European countries (France, Germany, Italy, Spain, and United Kingdom).

After obtaining the networks, several geometrical and topological measurements (e.g.~\cite{costa2007characterization}) are calculated, namely the average vertex degree, the standard deviation of the vertex degree, the standard deviation of the local vertex transitivity, the dispersion of the vertex position and the accessibility, which are considered as features characterizing each considered city.   After these features are collected and standardized, the coincidence method is applied, yielding respective networks, whose overall connectivity can be conveniently controlled by the parameter $\alpha$.  Noticeable results are obtained and discussed, including the organization of cities in well-defined groups sharing respective properties.  Interestingly, four main groups of cities have been obtained presenting the majority of cities from respective countries.  

To complement our study, we then apply the coincidence methodology on the weight matrices respective to each obtained network, so that the whole set of weights is understood as corresponding to the features characterizing each considered network.  The application of the coincidence method then allows a features network to be obtained characterizing in an accurate and objective manner the effect of the distinct possible feature combinations on the obtained networks topology.  The therefore obtained network was characterized by a well-defined modular structure, from which four respective communities were then identified, which can be understood as the four main models that can be obtained for the cities while considering different feature combinations.  In particular, it is suggested that the hubs of each of these communities can be understood as the respective prototype, therefore summarizing each of the four models in terms of a respective reference network.
It has also been found that the four obtained models share three of the five adopted features.

The present work starts by presenting how the data was obtained and then proceeds to describing the basic employed concepts and methods.  The results are presented next regarding the cities networks, and then to the features network.

\section{Materials and Methods}
\label{sec:materials}

The cities considered in this work are from five European countries (namely France, Germany, Italy, Spain and the United Kingdom) and have populations ranging from 200,000 and 300,000 inhabitants. 
Other than that, the cities were chosen arbitrarily. Four cities from each country were considered, totalling $n=20$ cities. Each city is represented by a \emph{streets network}, with nodes representing street intersections, while the edges joining two nodes correspond to the streets and avenues. The graphs were obtained from OpenStreetMap~\footnote{https://www.openstreetmap.org/}, a public and crowd-sourced repository of geographical information.

\begin{table}[ht]
   \caption{List of the analyzed cities, grouped by country, with corresponding populations. 
   Adapted from~\cite{united2019world}.}
   \centering
   \begin{tabular}{llc}
   \toprule
      City & Country & Population\\
   \midrule
Nantes & France & 280000 \\
Bordeaux & France & 230000 \\
Lille & France & 230000 \\
Rennes & France & 210000 \\
Brunswick & Germany & 240000 \\
Freiburg & Germany & 220000 \\
Kiel & Germany & 230000 \\
Augsburg & Germany & 260000 \\
Bari & Italy & 280000 \\
Messina & Italy & 220000 \\
Verona & Italy & 220000 \\
Padova & Italy & 200000 \\
Vigo & Spain & 300000 \\
Granada & Spain & 230000 \\
Oviedo & Spain & 220000 \\
Mostoles & Spain & 210000 \\
Bradford & U.K. & 300000 \\
Derby & U.K. & 270000 \\
Luton & U.K. & 260000 \\
Southampton & U.K. & 240000 \\
\bottomrule
   \label{tab:cities}
   \end{tabular}
\end{table}

The following five features have been considered in this work:

\begin{enumerate}
    \item \texttt{degmean}: the average of the vertex degrees;
    \item \texttt{degstd}: the standard deviation of the vertex degree;
    \item \texttt{transstd}: the standard deviation of the vertex local transitivity;
    \item \texttt{vposdisp}: dispersion of the point locations;
    \item \texttt{accessib}: the standard deviation of the vertex accessibility.
\end{enumerate}

The vertex degree distribution is a simple yet rich graph measurement, providing information about the connectivity of the nodes~\cite{costa2007characterization}. We consider two statistics from this distribution: the average and the standard deviation (\texttt{degmean} and \texttt{degstd}).

\begin{figure}[ht]
    \centering
    \includegraphics[width=.9\textwidth]{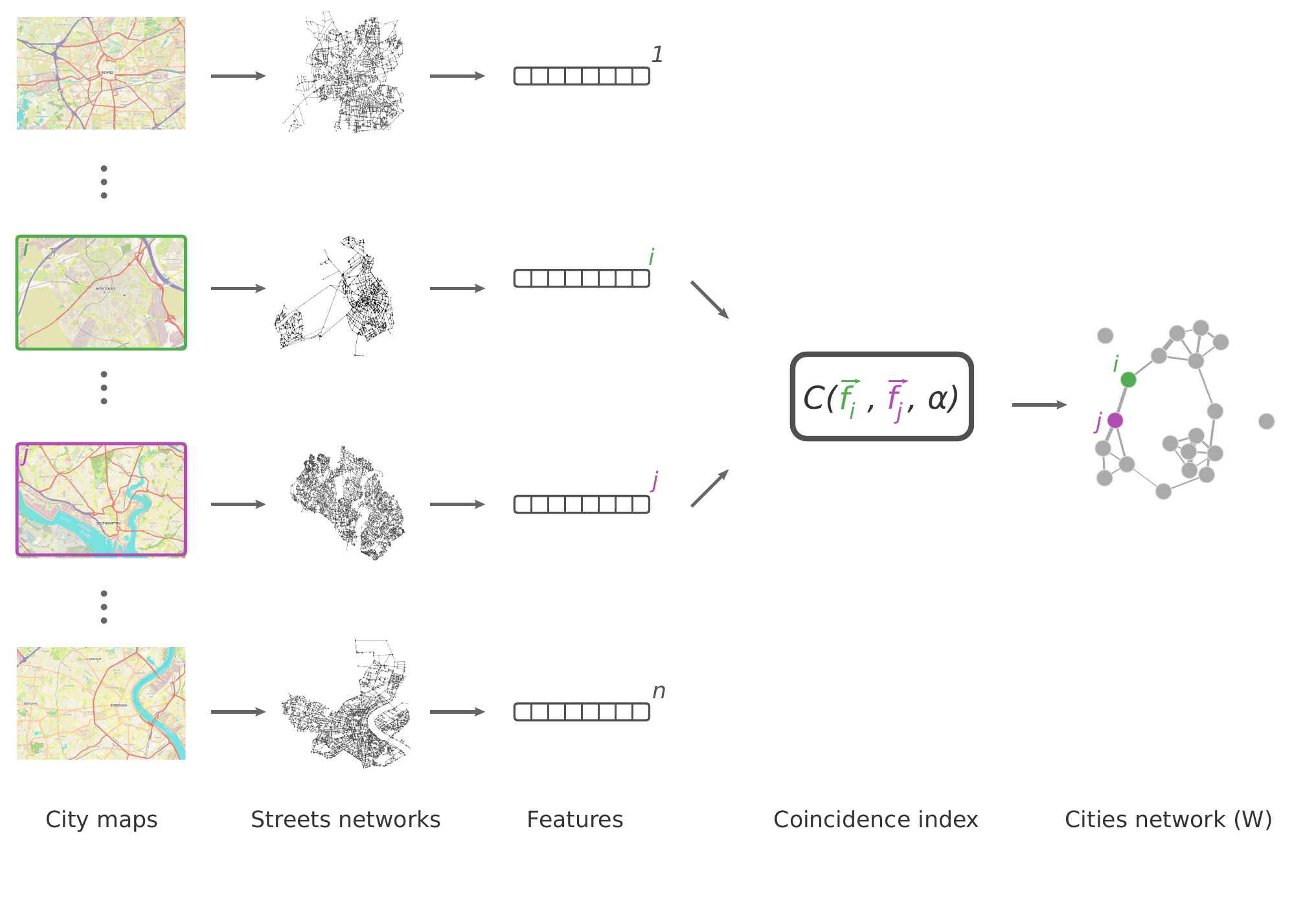}
    \caption{Diagram indicating how the cities networks is obtained.  The $n=20$ original
    cities, represented as maps here, are converted into respective street networks where
    each node corresponds to crossings of two or more streets while the links
    stand for the streets themselves.  Features, corresponding to five topological measurements,
    are then obtained for each of the street networks.  The similarity between the cities
    is then inferred by using the coincidence index, leading to the respective cities network.  }
    \label{fig:framework1}
\end{figure}

Fig.~\ref{fig:framework1} depicts the main steps adopted for obtaining the cities network.
It starts by representing each city in terms of a respective street network, in which the nodes
correspond to crossings of two or more streets, while the links stand for respective
streets.  Network measurements (topological and geometric) are obtained for each of the street network, and the similarity between each possible pair of cities is estimated in terms of the respective coincidence index.  The cities network is obtained by thresholding the coincidence values.

\begin{figure}[ht]
    \centering
    \includegraphics[width=.9\textwidth]{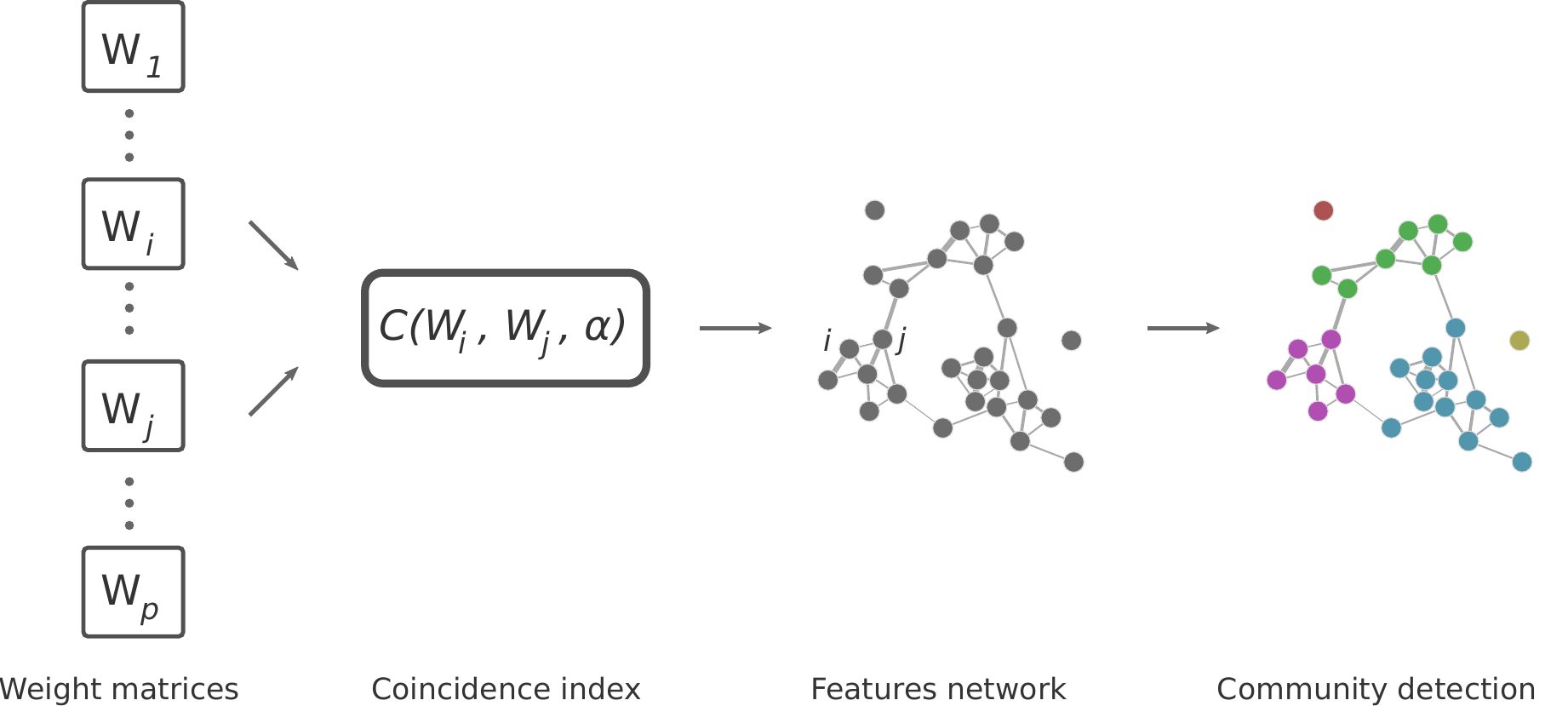}
    \caption{Diagram illustrating the estimation of the features network.  Weight matrices
    are obtained, by using the coincidence methodology, for each of the possible feature
    combinations ($p=31$).  The coincidence index is then calculated for each pair of weight
    matrices, yielding the features network, in which each node correspond to the network
    with respective feature combination.  The communities of the features network can 
    then be detected and understood as possible models of the original data.  In particular,
    the hub of each of the detected communities can be understood as respective \emph{prototypes}.}
    \label{fig:framework2}
\end{figure}

Fig.~\ref{fig:framework2} illustrates the estimation of the features network, starting
from the weight matrices $W_i$ corresponding to the networks obtained for each possible combination $i$ ($i=1 \ldots p=31$) of the adopted features.  The coincidence method is then applied in order to quantify the similarity between each of these matrices, yielding the \emph{features network}.  Each node in the latter network corresponds to a specific feature combination, while the links between
these nodes reflect the respective pairwise similarity.  Community detection can then be 
applied on the obtained features networks in order to identify the possible models respective
to the original data.

Another important characteristic of graphs is that the interconnectivity around each node can vary significantly.  One popular respective measurement is the \emph{clustering coefficient} (or transitivity)~\cite{costa2007characterization}. In this work, the standard deviation of the vertex transitivity is considered.

The geographical distribution of vertices is also a factor of (dis) similarity among city graphs, so that the dispersion of the vertex positions is in the current work taken as a network feature. It is computed by Eq.~\ref{eq:vposdisp}, where $\vec{p_i}$ represent the vertex position for each vertex, $m$ is the number of vertices and $\vec{c}$ is the centroid of the positions. As the regular standard deviation, it measures the dispersion of the data around the mean. Here we considered $||\cdot||$ the $L_2$ norm.

\begin{equation}
    vposdisp = \frac{\sum_i^n ||\vec{p_i} - \vec{c}||}{m}
    \label{eq:vposdisp}
\end{equation}

Another feature that has been successfully employed for city analysis is the graph accessibility~\cite{travenccolo2008accessibility,de2014role,viana2013accessibility}, which can be understood as a generalization of the concept of vertex degree. The vertex accessibility quantifies its influence over the respective neighbourhood. This measurement assumes a dynamics of particular interest, which in this case is the self-avoiding random walk. A parameter $h$ defines the order of the neighbourhood (e.g.~$h=2$ indicates the nodes that are at topological distance 2 from the reference node).

Each of the 20 considered cities was characterized in terms of the above discussed measurements.  In order to ensure more commensurate values between these five features, they are
respectively standardized (e.g.~\cite{costa2021coincidence}) so as to have null mean and unit variance.  The standardization of each of the features $f_i$ can be performed as:
\begin{equation}
    \tilde{f}_i = \frac{f_i - \mu_{f_i}} {\sigma_{f_i}}
\end{equation}

where $\mu_{f_i}$ and $\sigma_{f_i}$ correspond to the mean and standard deviation of $f_i$, respectively.

The \emph{coincidence similarity} between two real-valued feature vectors $\vec{f}$
and $\vec{g}$ can be expressed~\cite{costa2021onsimilarity,costa2021coincidence} as:
\begin{equation}
   \mathcal{C}_R(\vec{f},\vec{g}) = \mathcal{I}_R(\vec{f},\vec{g}) \  \mathcal{J}_R(\vec{f},\vec{g})
\end{equation}

where $\mathcal{I}_R(\vec{f},\vec{g})$ is the \emph{interiority index} between  $\vec{f}$
and $\vec{g}$, calculated as:
\begin{equation}
   \mathcal{I}_R(\vec{f},\vec{g}) = \frac{ \sum_i \min\left\{ | f_i |, | g_i |\right\}}
   {\min\left\{ \sum_i | f_i |, \sum_i | g_i | \right\}}
\end{equation}

and  $\mathcal{J}_R(\vec{f},\vec{g})$ is the \emph{Jaccard index} between the two
real-valued vectors, given as:
\begin{equation}
   \mathcal{J}_R(\vec{f},\vec{g}) = \frac{ \sum_i sign(f_i \ g_i) \min\left\{ | f_i |, | g_i |\right\}}
   {\sum_i \max\left\{ | f_i |, | g_i |\right\}}
\end{equation}

It is of particular interest to adopt the $\alpha$ ($0 \leq \alpha \leq 1$) 
parameter~\cite{costa2021onsimilarity,costa2021coincidence}  as a means to control the contributions of the pairwise aligned and anti-aligned signs of the involved feature values on the overall
coincidence result.  This can be immediately implemented~\cite{costa2021onsimilarity,costa2021coincidence} 
as:
\begin{equation}
   \mathcal{C}_R(\vec{f},\vec{g},\alpha) = \mathcal{I}_R(\vec{f},\vec{g}) \  \mathcal{J}_R(\vec{f},\vec{g},\alpha)
\end{equation}

where:
\begin{equation}
   \mathcal{J}_R(\vec{f},\vec{g},\alpha) =
    \frac{\sum_i \alpha   |s_{f_i} + s_{g_i}| \min\left\{ | f_i |, | g_i |\right\} 
   - (1 - \alpha)  |s_{f_i} - s_{g_i}| \min\left\{ | f_i |, | g_i |\right\} }
   {\sum_i \max\left\{ | f_i |, | g_i |\right\}}
\end{equation}

where $s_{f_i} = sign(f_i)$.  
We also have that $-2(1-\alpha) \leq J_R(\vec{x}, \vec{y}, \alpha) \leq 2\alpha$.

When $\alpha=0.5$ we have that $\mathcal{J}_R(\vec{f},\vec{g},\alpha) = \mathcal{J}_R(\vec{f},\vec{g})$. 
For $\alpha>0.5$, the pairwise features having the same sign will have greater
contribution than those with opposite signs.  The contrary effect is obtained
when $\alpha<0.5$.  As a consequence, the overall degree of interconnection 
between the nodes of the resulting networks can be controlled by varying
$\alpha$, with more interconnected structures being obtained for larger
values of $\alpha$.  It has been observed that $\alpha <0.5$ tends to significantly
enhance the modularity and levels of interconnectivity detail of the respectively obtained 
networks~\cite{costa2021onsimilarity,costa2021coincidence}.

\section{Topological Study of European Cities}

\begin{figure}[ht]
    \centering
    \includegraphics[width=.9\textwidth]{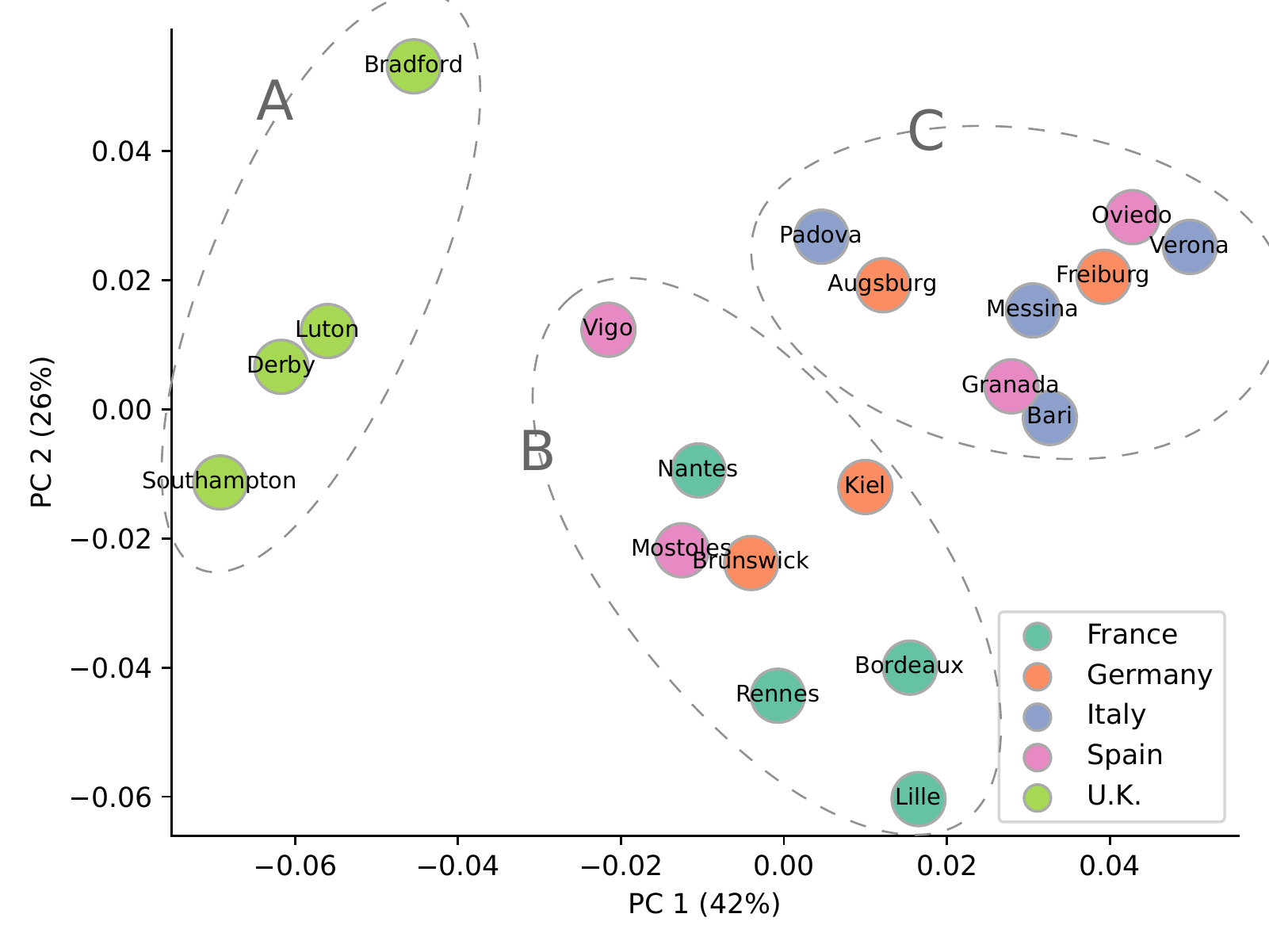}
    \caption{Principal component analysis of all (5) features considered. The axes shown in the figure correspond to the first two principal components.  The percentage of variance
    explained by each axis is shown respectively. }
    \label{fig:pca}
\end{figure}

As a first step, aimed at obtaining a preliminary comparison reference for our studies,
we obtained the principal component analysis (PCA, e.g.~\cite{gewers2021principal}) of the 20 cities, which is shown in Fig.\ref{fig:pca}, while taking into account all the respective five measurements. The total variance explanation accounted by the first two principal axes is relatively low ($68\%$), indicating that the adopted measurements are little correlated one another, therefore effectively complementing the characterization of the city structures. The resulting distribution of the cities in the obtained PCA is not uniform, with a concentration being observed on the right-hand side.  Three main groups or clusters can be observed (A, B, and C).  Group A contains 4 British cities. Group B contains cities from Spain, Germany, and France, while Group C includes cities Spain, Germany and Italy. Groups B and C are, therefore, largely heterogeneous.  

The transformation of the cities dataset into respective networks involves only two parameters: the overall threshold $T$, and the parameter $0 \leq \alpha \leq 1$ controlling the relative contribution of aligned and anti-aligned features signs (two types of joint variations).  By testing several combinations of these two parameters, we identified that $T=0.10$ constitutes a particularly adequate choice in the sense of yielding modular structures and enhancing the interconnectivity details for several values of $\alpha$.   As discussed in Sec.~\ref{sec:materials}, the parameter $\alpha$ can be used to control the overall level of interconnections between the nodes in the obtained networks, in the sense that the larger the value of $\alpha$, the more connected the obtained networks will be.  Interestingly, the adoption of smaller values of $\alpha$ generally contributes to enhancing substantially the overall modularity and details in the obtained networks~\cite{costa2021coincidence}.  

Another important characteristic observed from the adopted multi-alpha analysis is that
the connections established for a given $\alpha$ will be necessarily preserved for larger
values of that parameter.  Therefore, `early' connections obtained for a relatively small value $\alpha_1$ can be understood as being stronger and more stable in more interconnected networks
obtained for larger values of $\alpha > \alpha_1$.

\begin{figure}[ht]
    \centering
    \begin{subfigure}[b]{.32\textwidth}
        \centering
        \includegraphics[width=\textwidth]{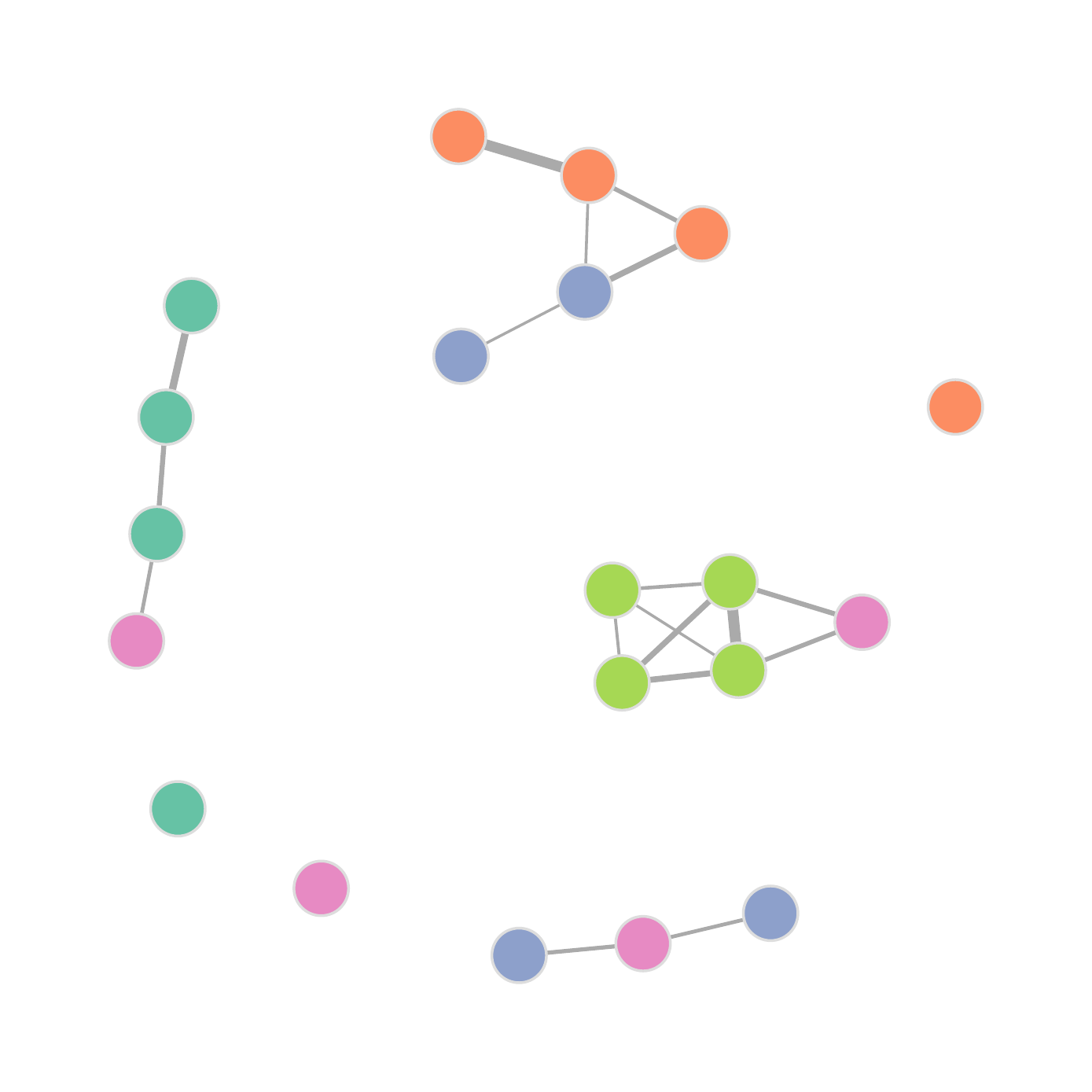}
        \caption{$\alpha=0.25$}
    \end{subfigure}
    \begin{subfigure}[b]{.32\textwidth}
        \centering
        \includegraphics[width=\textwidth]{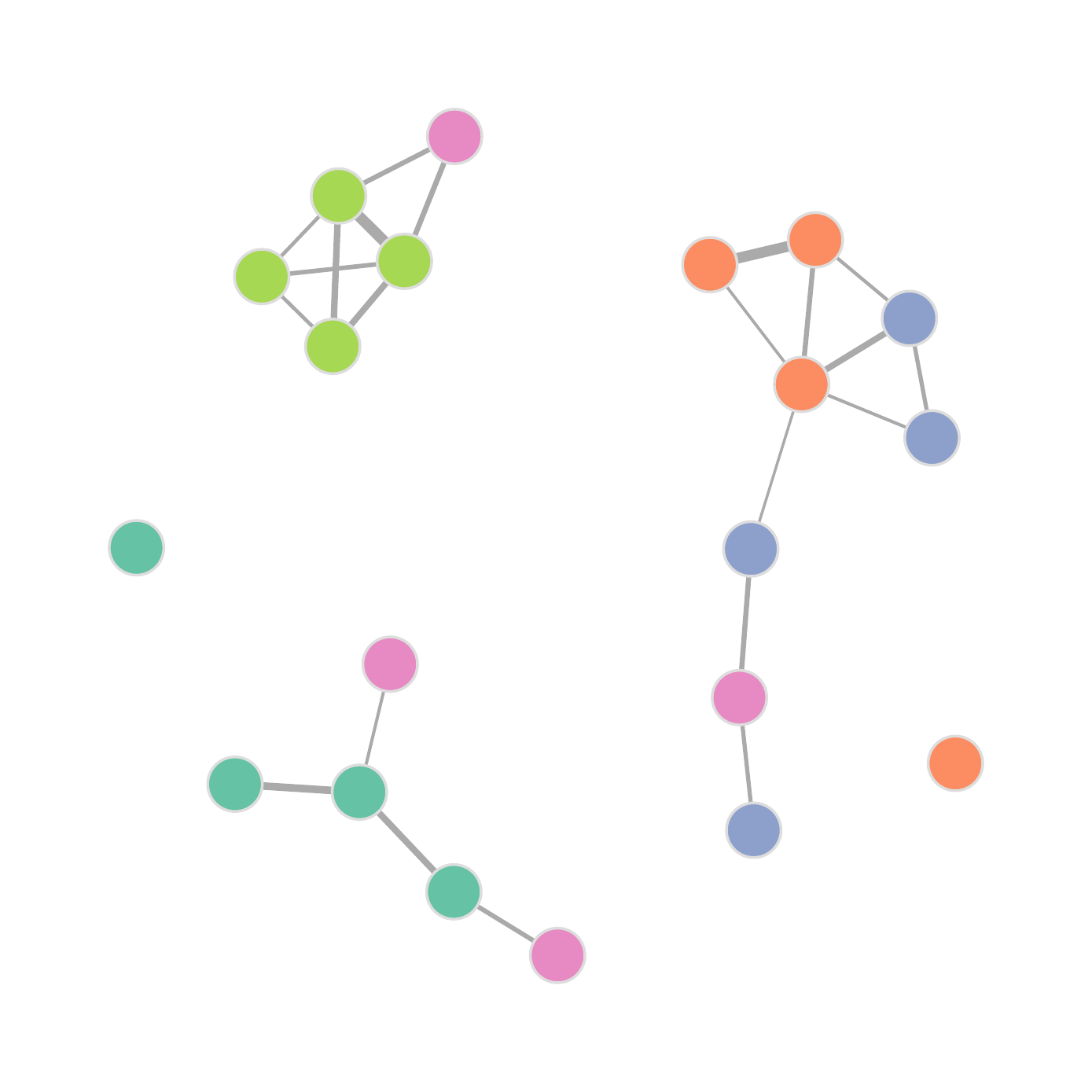}
        \caption{$\alpha=0.32$}
    \end{subfigure}
    \begin{subfigure}[b]{.32\textwidth}
        \centering
        \includegraphics[width=\textwidth]{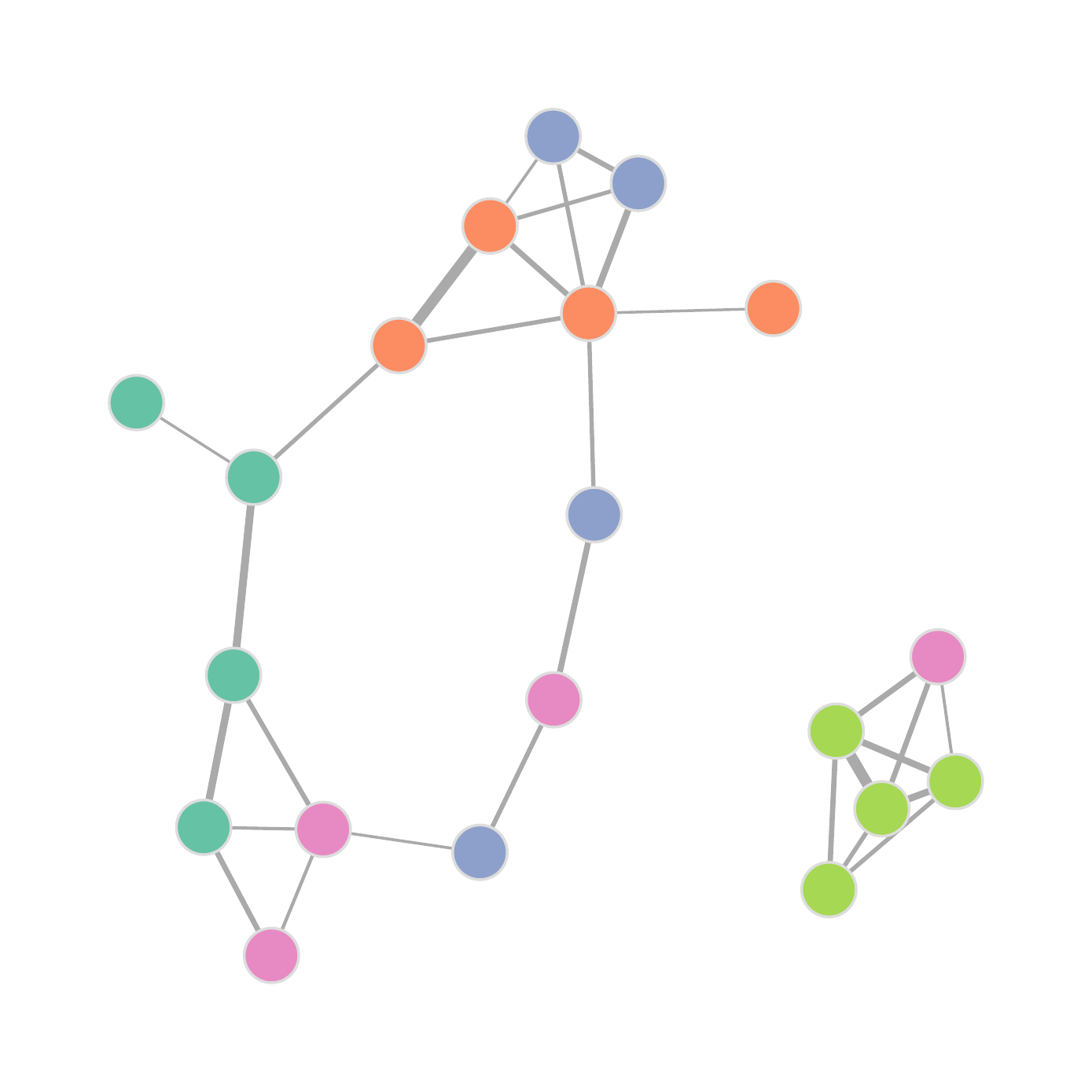}
        \caption{$\alpha=0.39$}
    \end{subfigure} \\
    \begin{subfigure}[b]{.32\textwidth}
        \centering
        \includegraphics[width=\textwidth]{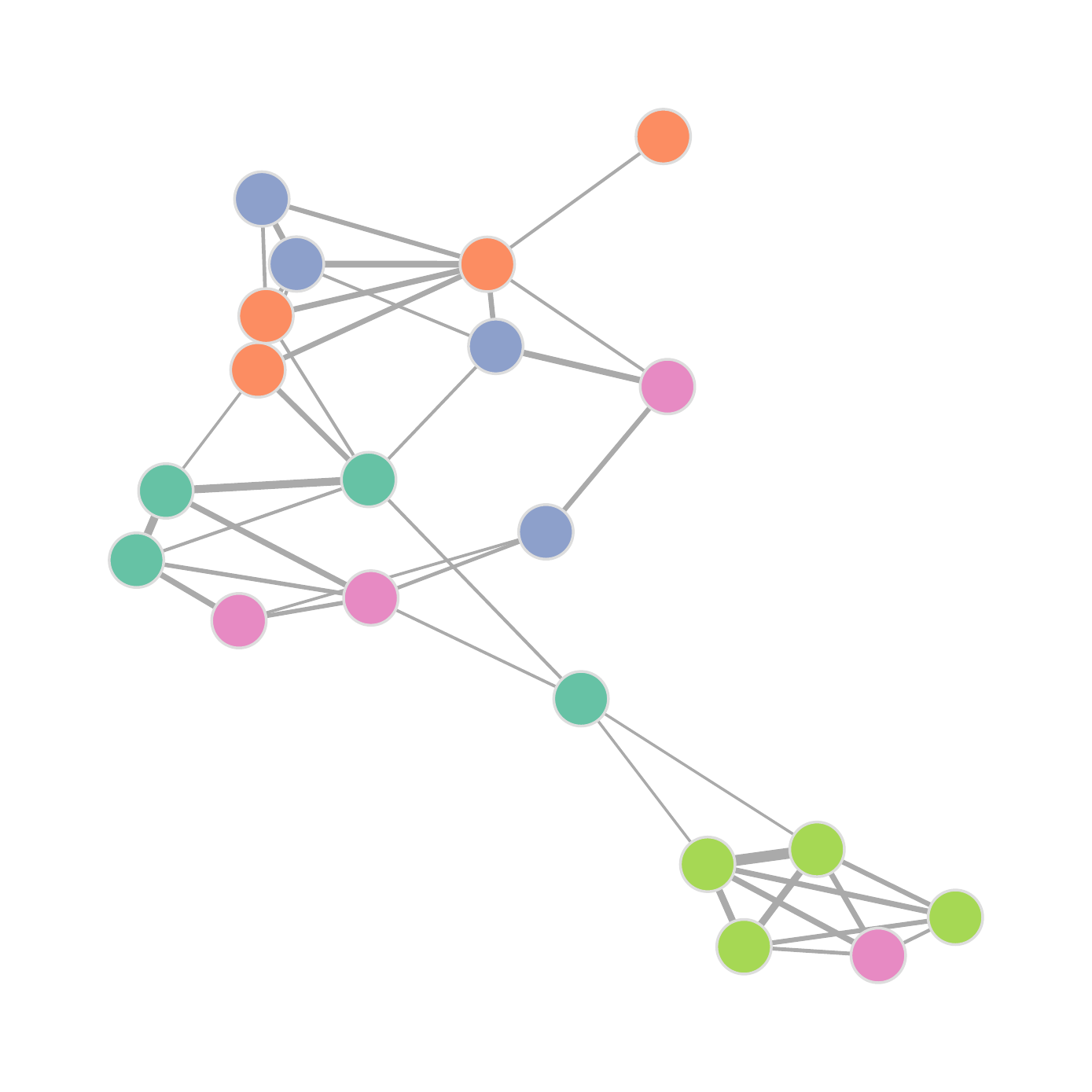}
        \caption{$\alpha=0.46$}
    \end{subfigure}
    \begin{subfigure}[b]{.32\textwidth}
        \centering
        \includegraphics[width=\textwidth]{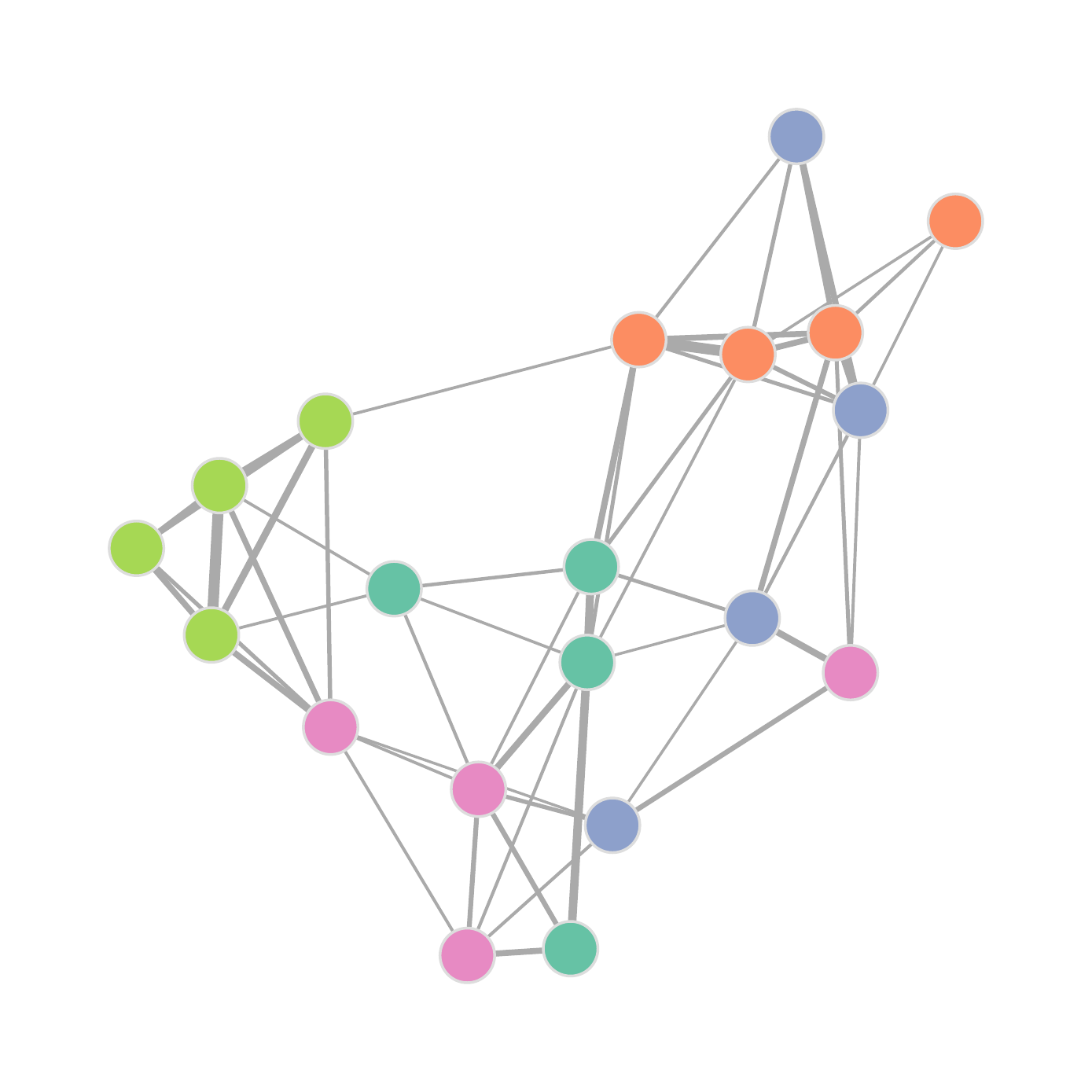}
        \caption{$\alpha=0.53$}
    \end{subfigure}
    \begin{subfigure}[b]{.32\textwidth}
        \centering
        \includegraphics[width=\textwidth]{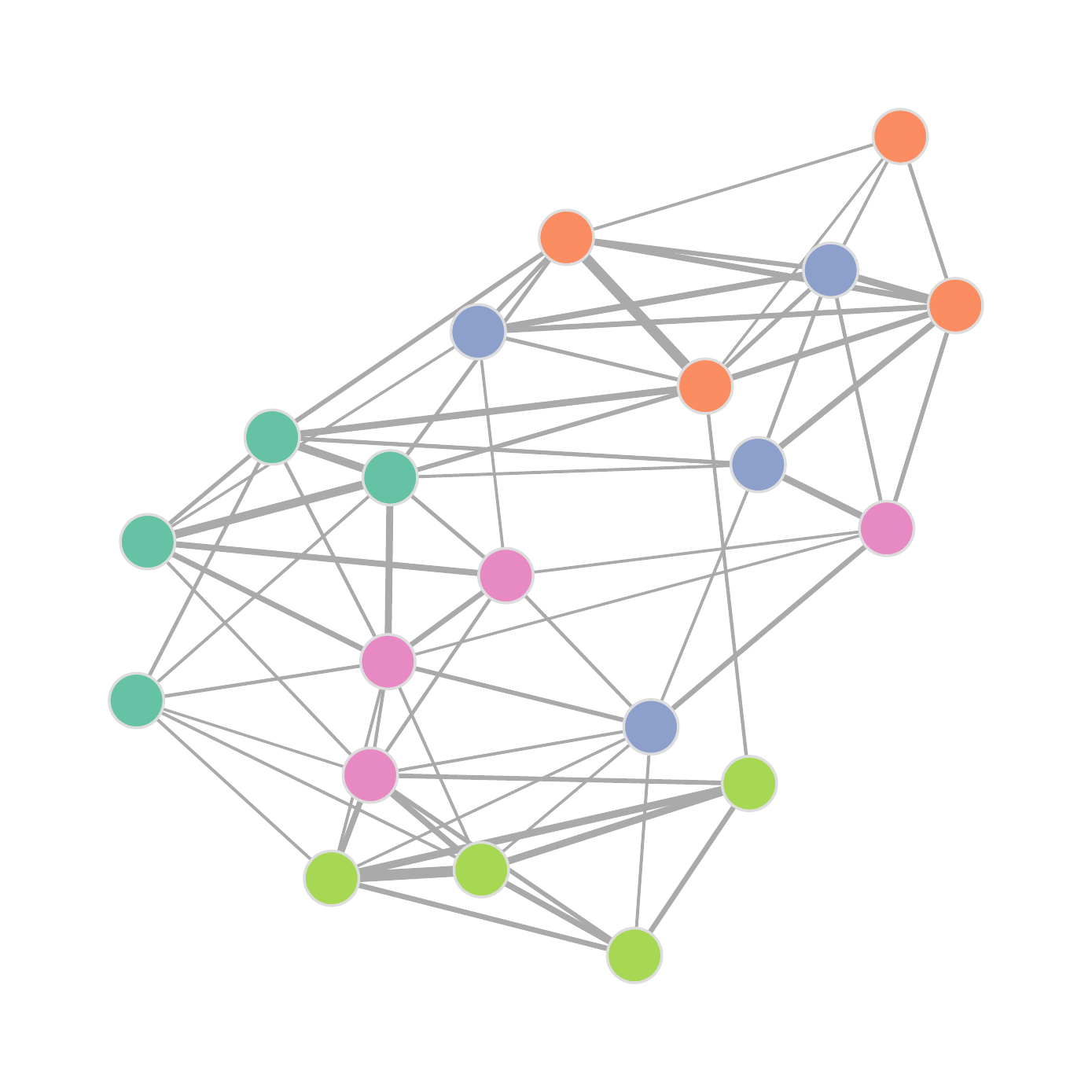}
        \caption{$\alpha=0.60$}
    \end{subfigure} 
    \caption{Cities networks obtained for different values of $\alpha$ by the coincidence index calculation. The same edge threshold of $T=0.10$ was considered for all graphs.  As expected, the overall
    connectivity tends to increase with $\alpha$.  Particularly detailed and modular networks
    were obtained for the smaller values of $\alpha$.}
    \label{fig:alphas}
\end{figure}

Fig.~\ref{fig:alphas} shows the cities networks obtained for $T=0.10$ and
$\alpha =$ 0.25, 0.32, 0.39, 0.46, 0.53, 0.6.   The five node colors identify the respective countries to
which the cities belong.   As expected, little connected networks have been obtained
for the two smallest values of $\alpha$ (a),(b).  However, these two networks also contain four
connected components presenting a manifest homogeneity, in the sense of including
cities mostly from the same country, as can be appreciated from the colors within each component.

As $\alpha$ is increased (Fig.~\ref{fig:alphas}c), three of the four groups in (b) merge 
into a major community, with the remainder group corresponding mostly to the
British cities.  After increasing $\alpha$ further (d), all groups coalesce into a single
component, with the community containing mostly British cities 
connected through a single link to the remainder of the network.  As $\alpha$ is then increased to obtain the networks in (e) and (f), the respective networks
become more and more interconnected at the expense of the respective modularity
and level of details, which are both substantially decreased.  Even so, many neighboring
cities in (e) and (f) tend to be from the same country.

It is also interesting to keep in mind that substantially more information and
insights can be obtained by considering the multi-alpha analysis such as that shown
in Fig.~\ref{fig:alphas}, than by considering a single value of $\alpha$.  Indeed, in addition to
the already observed possibility of identifying the strongest links appearing
for the first values of $\alpha$, the multi-alpha analysis also plainly indicates
how the initial modules progressively merge while their relative adjacencies
are mostly maintained as $\alpha$ increases.  

The obtained results corroborate the critical role of the parameter $\alpha$ respectively
to the obtained highly modular networks with detailed interconnectivity.  Indeed, had only
the standard configuration of $\alpha=0.5$ been used, the obtained result would
be characterized by almost no modularity and little level of details.  It was only thanks
to the possibility to vary $\alpha$ to smaller values that the identification of highly modular and detailed networks have become possible. 

\begin{figure}[ht]
    \centering
    \includegraphics[width=0.5\textwidth]{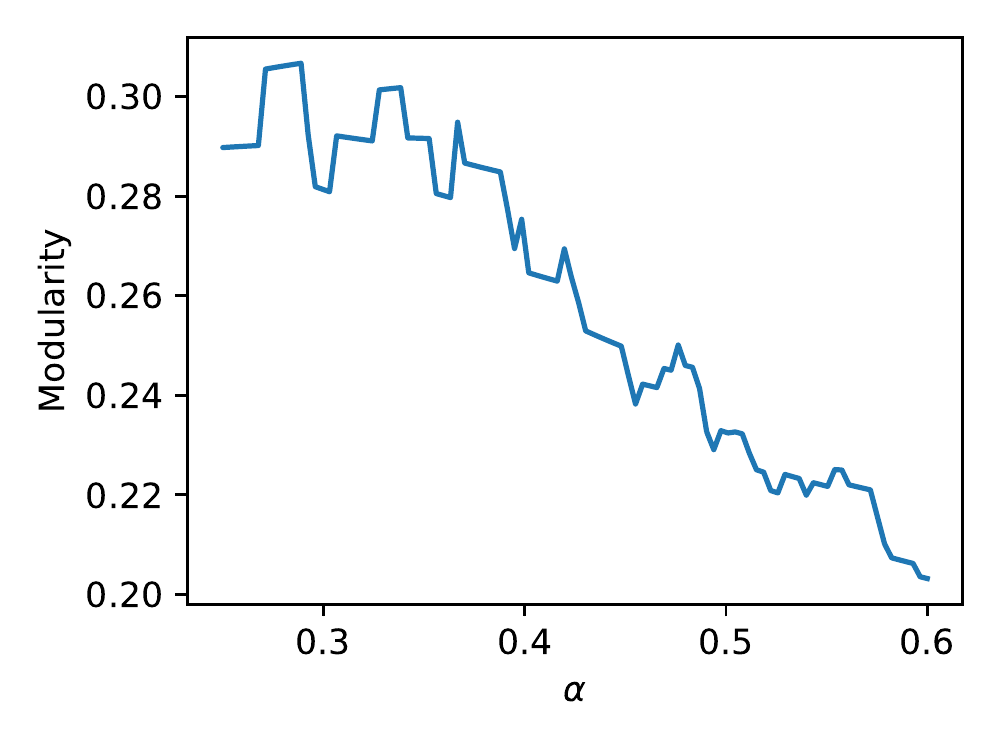} \quad
    \caption{Modularity value in terms of the parameter $\alpha$, respectively 
    to city country membership. Observe that the modularity 
    tends to decrease with $\alpha$.  Its maximum value was found to occur at $0.29$.}
    \label{fig:modularity}
\end{figure}

A particularly remarkable result observed in Fig.~\ref{fig:alphas} is the noticeable uniformity
of the communities obtained for small values of $\alpha$ resulting from networks (a) and (b).
This result indicates that the adopted features and network construction methodology
were accurate enough to reveal impressive levels of topological homogeneity between
cities from a same country.   This important result motivated us to proceed further in the
sense of trying to identify the value of $\alpha$ leading to the greatest possible modularity
given the adopted features.  In order to do so, we considered several values of $\alpha$ between
0.25 and 0.6 (with resolution $0.07$) and calculated the respective modularity (e.g.~\cite{fortunato2010community}), so that its maximum
could be identified.  The modularity was calculated by using the country membership
as reference, in the sense that the maximum modularity would correspond to obtaining
completely homogeneous communities.  Fig.~\ref{fig:modularity} depicts the modularity obtained for the
considered cities and features in terms of the parameter $\alpha$.   

\begin{figure}[ht]
    \centering
    \includegraphics[width=.7\textwidth]{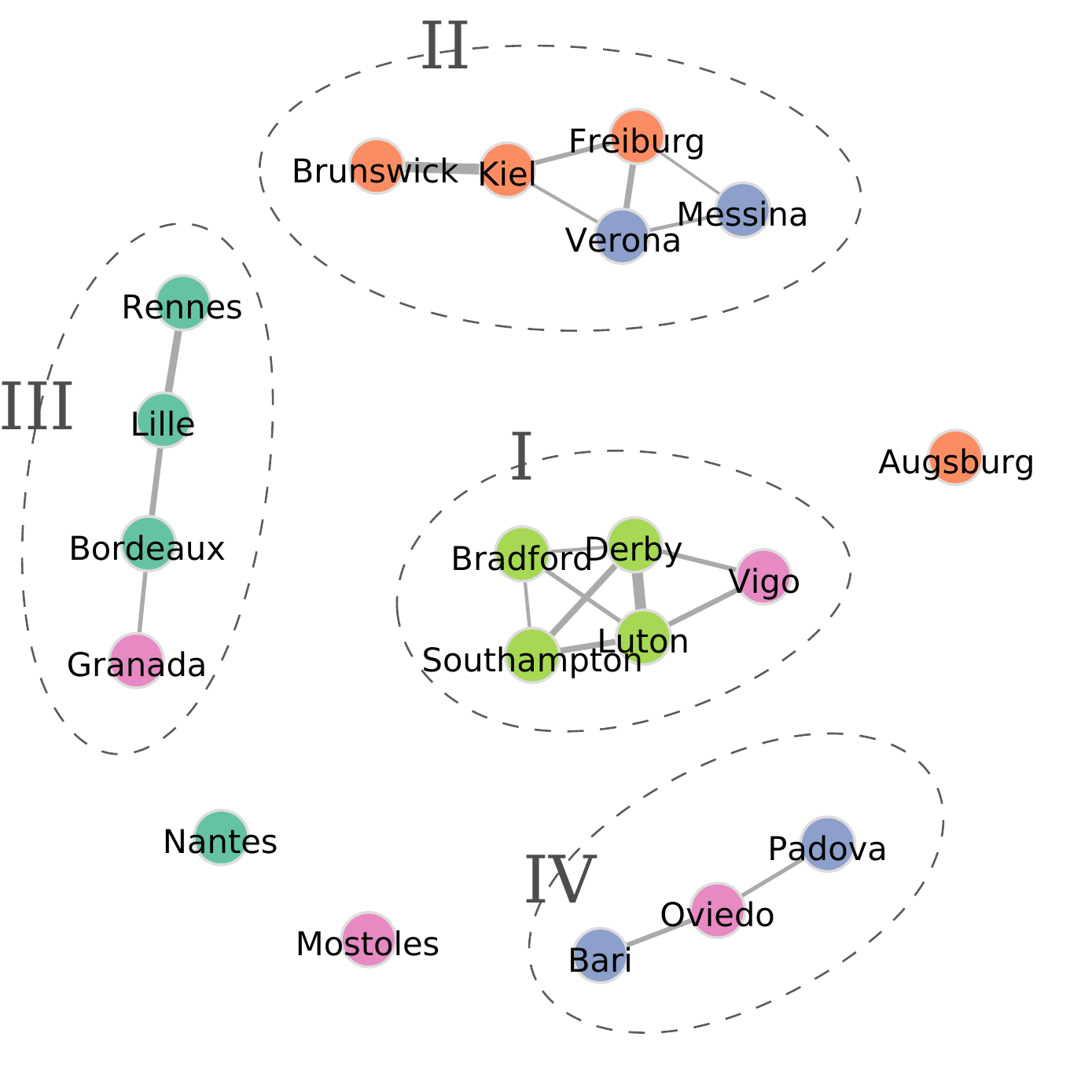} 
    \includegraphics[width=.15\textwidth]{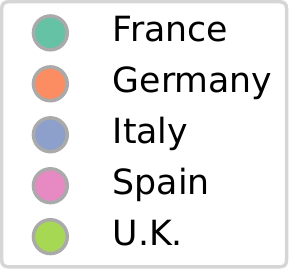} 
    \caption{The cities network obtained by using the coincidence
    methodology with $T=0.10$ and $\alpha=0.29$, which implied a modularity of $0.31$. The nodes represent the cities and the connections widths are proportional to the values of the coincidence index between the 
    features of the two respective cities. Four separated components have been
    obtained, which can be mostly associated to the
    United Kingdom (I), Germany (II), France (III) and Italy (IV).}
    \label{fig:coincgraph}
\end{figure}

As could be expected from our previous experiment involving 6 values of $\alpha$, the
modularity decreases substantially for the larger values of $\alpha$, with a respective
maximum being observed at $\alpha_M = 0.29$.   The respectively defined network, characterized by the maximum possible overall modularity for the considered cities and features,
is shown in Fig.~\ref{fig:coincgraph}.

\begin{figure}[ht]
    \centering
    \begin{subfigure}[b]{0.8\textwidth}
    \includegraphics[width=\textwidth]{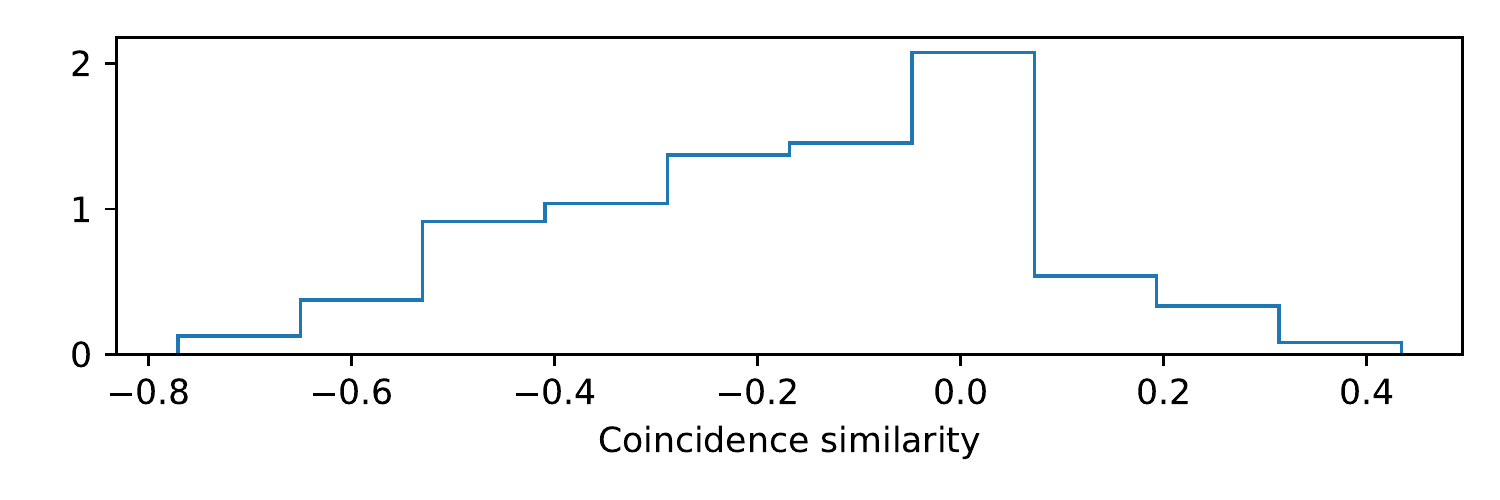}
    \caption{}
    \end{subfigure}\\
    \begin{subfigure}[b]{0.8\textwidth}
    \includegraphics[width=\textwidth]{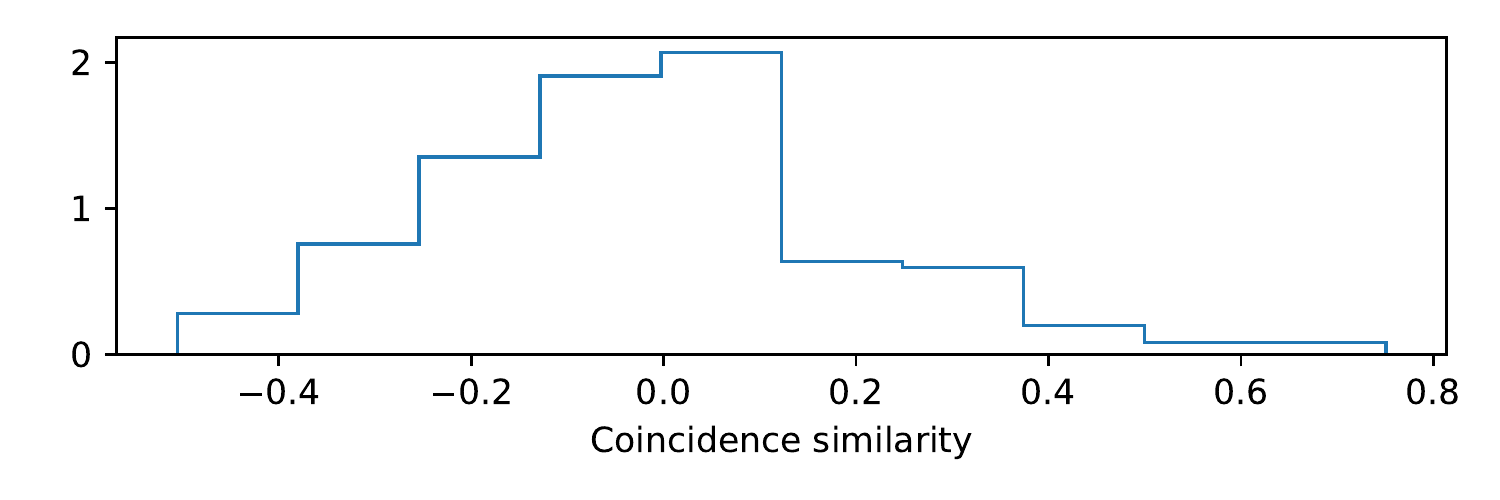} 
    \caption{}
    \end{subfigure}
    \caption{Relative frequency histograms of coincidence similarity values
    obtained respectively to the maximally modular cities network (a) and for $\alpha=0.5$ (b), with respective average $\pm$ standard deviation values of $-0.16 \pm 0.23$ and $-0.019 \pm 0.21$.}
    \label{fig:histcoinc}
\end{figure}

Fig.~\ref{fig:histcoinc}(a) shows the relative frequency histogram of 
the coincidence
similarity values obtained for the maximally modular network ($\alpha_M = 0.29$).
Given that $\alpha <0.5$, implying the positive joint feature signs to be penalized,
the obtained average becomes negative which, at least for this case, results in a more
detailed network.  The relative frequency histogram obtained for $\alpha=0.5$,
shown in Fig.~\ref{fig:histcoinc}(b), has mean close to zero and standard
deviation comparable to that obtained for the maximally modular network, 
therefore indicating the presence of higher coincidence values, which tends to 
yield a more densely connected network with less details and smaller modularity.
Observe also the markedly distinct shapes of the two
obtained relative frequency histograms, implied by the two distinct values of $\alpha$.  This corroborates the important fact that the variation of the parameter $\alpha$ has an effect that goes beyond transforming the coincidence values in a trivial manner (e.g.~shifting or scaling).

The maximally modular network in Fig.~\ref{fig:coincgraph} can now be compared to the 
PCA in Fig.~\ref{fig:pca}.  Other than the group A containing only British cities,
the marked modularity obtained by the coincidence method cannot be directly inferred from 
PCA results, which substantiates the potential of the reported methodology for
characterizing the interrelationship between the data elements (cities).   Indeed, though the adopted PCA approach considers all the five original features, it is intrinsically limited to being a two-dimensional projection of the original data elements that cannot preserve all the original information.  Conversely, the links in the obtained maximally modular network take into account all the five original features while  not involving any projection or information loss other than those implied by the adopted (optional) thresholding.

The  network in  Fig.~\ref{fig:coincgraph}  presents four major connected components, which have
been labelled as I, II, III and IV by decreasing interconnectivity (total strength).  
Four of the cities in group I are British, with the Spanish city of Vigo being also included.
The cities in group II are predominantly German, also encompassing two Italian cities.  
Three French cities, plus the Spanish city of Granada, compose group III.  The last group,
IV, contains two Italian and one Spanish city.  Three cities from distinct 
countries remained isolated for this value of $\alpha$: Augsburg, Nantes and Mostoles.   Therefore, these four obtained 
groups can be associated to respective countries as:  I $\leftrightarrow$ Britain; II $\leftrightarrow$ Germany; 
III $\leftrightarrow$ France, IV $\leftrightarrow$ Italy.  Interestingly, the adopted Spanish cities have not been associated to a respective group, which is likely a consequence of their distinct characteristics discussed in the following.  Mostoles is a municipality adjacent to Madrid which occupies a mostly plain terrain.  Though with ancient origins, it expanded mainly along the 20th century, therefore tending to present a more planned organization.  Vigo is a port city, therefore presenting a distinctive topology limited by the coastline.  Oviedo's central region is mostly elongated along a valley. The northeastern portion of Granada is hilly, having
nearly twice as much area as Oviedo.

It is also interesting to observe that the cities appearing in 
minority in the four main groups very probably present values of the adopted features that  make them indeed more similar to the groups to which they have been respectively assigned.

To conclude this section, it is interesting to make some additional considerations
regarding the remarkable obtained cities network presenting enhanced modularity and
uniformity regarding respective countries.  Indeed, there are several conditions
necessary for reaching this result, including an appropriate selection of features
capable of effectively describing the distinctive properties of the cities, as well as the
adoption of a sound and accurate methodology for translating from these features to
a respective network with high uniformity and homogeneity.  Plainly, these two
conditions have been mostly met in the reported approach: (i) the adopted five features,
despite their relatively small number, provided a specific characterization of the cities;
and (ii) the coincidence method confirmed its tendency to provide strict and
accurate characterization of the interrelationships between the original data
elements (cities), yielding a highly modular and detailed network representation
of the relationships between the considered cities.  

However, there is a third
condition for obtaining the remarkable results in Fig.~\ref{fig:coincgraph}, and this consists
in the fact that the adopted cities need to present, from the outset, distinctive
characteristics between cities of different countries while presenting 
homogeneity regarding a respective country.  The obtained results corroborate this
interesting possibility, suggesting that European cities (at least those considered here) tend to
present surprising homogeneity within a same country, while being relatively different between countries.  This interesting tendency may be related to
intrinsic climates, distinct planning traditions, as well as intrinsic socio-economic characteristics.

\section{Features Interrelationship}

It is well known (e.g.~\cite{da2010shape,han2011data}) that the features adopted
for characterizing data elements can greatly influence and even determine the results
of respective classifications.   This problem is so important that whole areas,
including \emph{feature selection} 
(e.g~\cite{chandrashekar2014survey,guyon2003introduction,li2017feature}), are to a large extent dedicated to studying
how the choice of different types of features can impact on pattern recognition and
machine learning.

Having obtained remarkable results while translating the 20 considered cities into
a respective modular and detailed network, it remains an equally interesting problem
to study to which an extent the adopted features contributed to the reported results.
Interestingly, it is possible to apply the same coincidence methodology employed to
obtain the networks also for this finality~\cite{costa2021elementary}.  More specifically,
networks are obtained respectively to several combinations of features, and the
coincidence method is them applied over the obtained weight matrices so as obtain
a new network in which each node corresponds to one of the feature combination network,
while the links correspond to the pairwise similarity between the weights of respective
matrices.  This interesting issue constitutes the main subject of the present 
section.

\begin{figure}[ht!]
    \centering
    \includegraphics[width=0.95\textwidth]{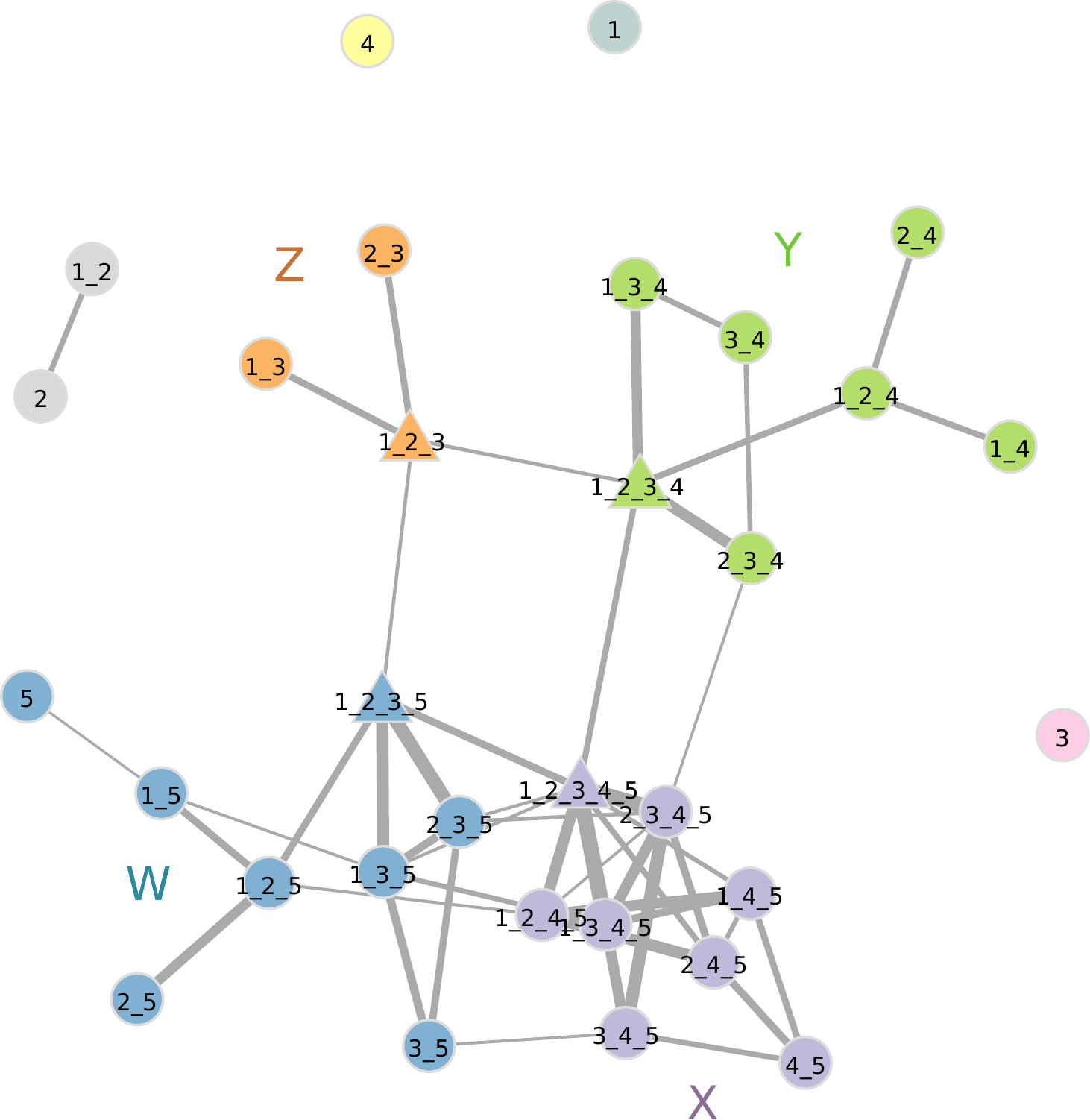}
    \caption{Network of feature combinations. Each vertex corresponds a configuration of features, while the edges reflect the coincidence index between the two coincidence graphs.  The width of the links are
    proportional to the respective pairwise coincidence values. The  communities were detected by using using the approach described in~\cite{blondel2008fast}.
    The hub within each identified community is shown as a triangle.}
    \label{fig:featnetwork}
\end{figure}

First, we obtained networks for each of the $p=31$ possible combinations  
($2^5$, except for the null combination) of the adopted
features.  Then, the coincidence method with $T=0.10$ and $\alpha=0.29$ was applied
considering the obtained weight matrices as features.  The thus obtained features network is shown
in Fig.~\ref{fig:featnetwork}

Interestingly, the resulting network, henceforth called \emph{features network} 
also resulted strongly modular (modularity equal to 0.448), with four well defined
communities (labelled W, X, Y, and Z) having been detected by the multilevel community finding method~\cite{blondel2008fast}.

In a similarity network with well-defined modularity, each of the nodes belonging to any of its
communities will tend to be similar to the other nodes in that same community.
Indeed, this corresponds to one of the possible rationales behind the very concept of
network modularity.   In this sense, each of the nodes in each of the obtained four 
communities in  Fig.~\ref{fig:coincgraph} corresponds to respectively similar cities.  This important property allows us to conclude that, in the
case of the 20 adopted cities, there are four possible main respective representations or \emph{models} of the
cities as networks, corresponding to each of the four modules, identified as
W, X, Y, and Z. 
Given that the hub within each of the detected communities is the node most
intensely interconnected to the others in the same community, it becomes possible
to consider that hub as a \emph{prototype} of the respective cities networks 
represented by that community.  The hubs identified respectively to each of the four 
communities in Fig.~\ref{fig:coincgraph} can be identified as having triangular shape.
Interestingly, the four obtained hubs can be found to be interconnected along a
square motif corresponding to the intersection between the four detected
communities.

\begin{figure}[ht!]
    \begin{subfigure}[b]{0.07\textwidth}
        \includegraphics[width=\textwidth]{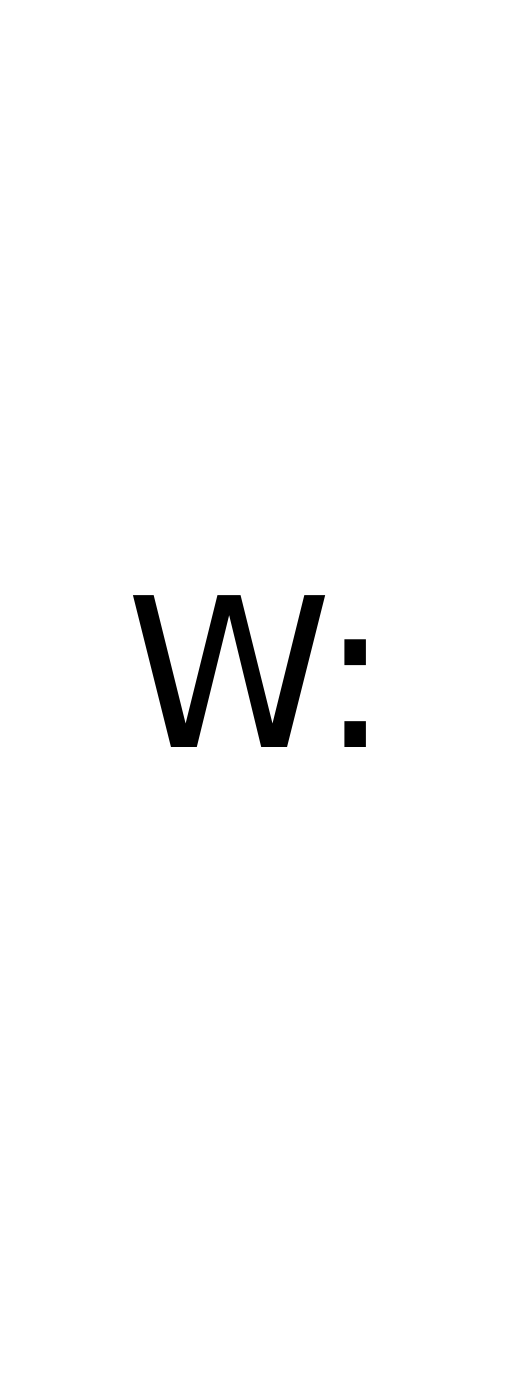}
    \end{subfigure}
    \begin{subfigure}[b]{0.2\textwidth}
        \includegraphics[width=\textwidth]{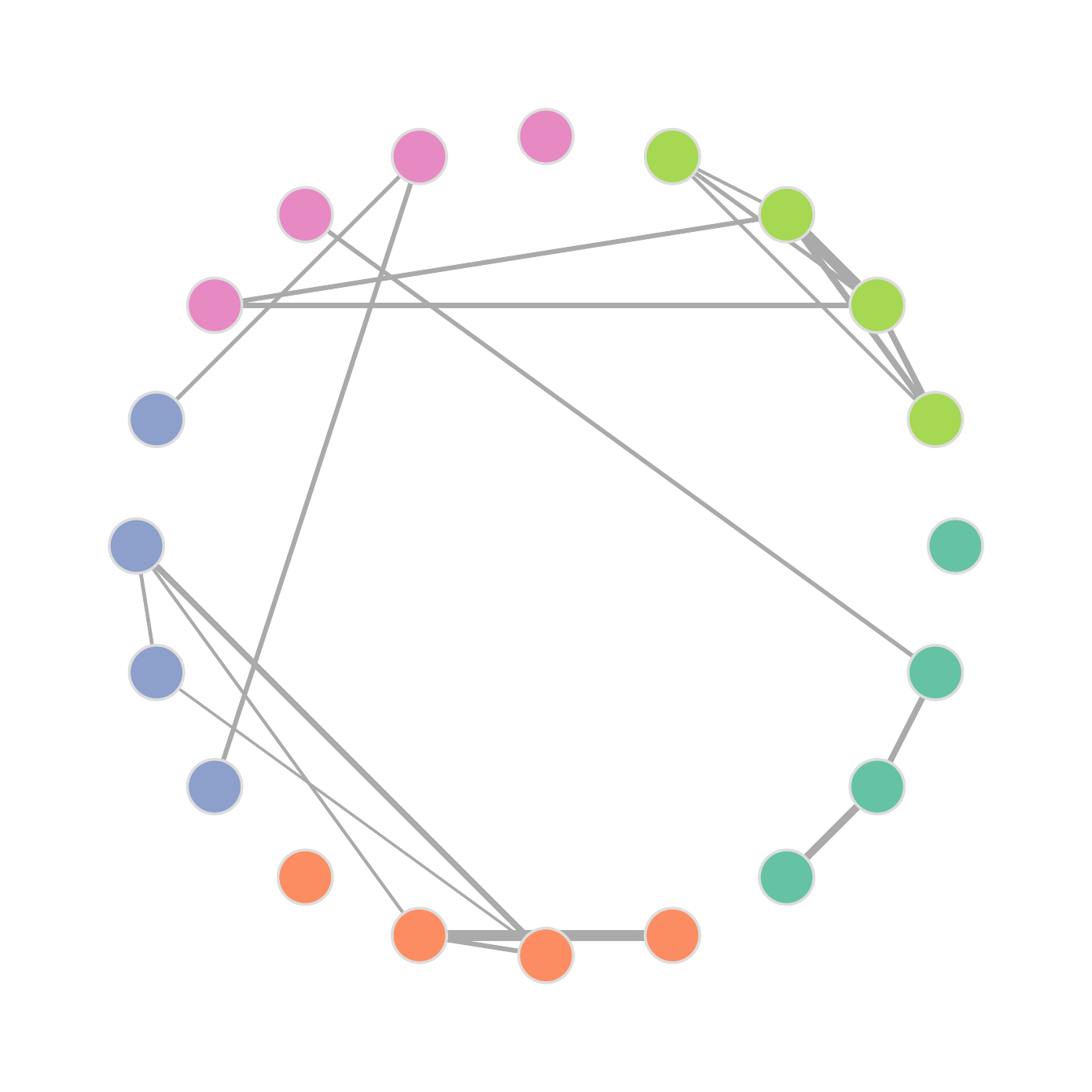}
    \end{subfigure}
    \begin{subfigure}[b]{0.2\textwidth}
        \includegraphics[width=\textwidth]{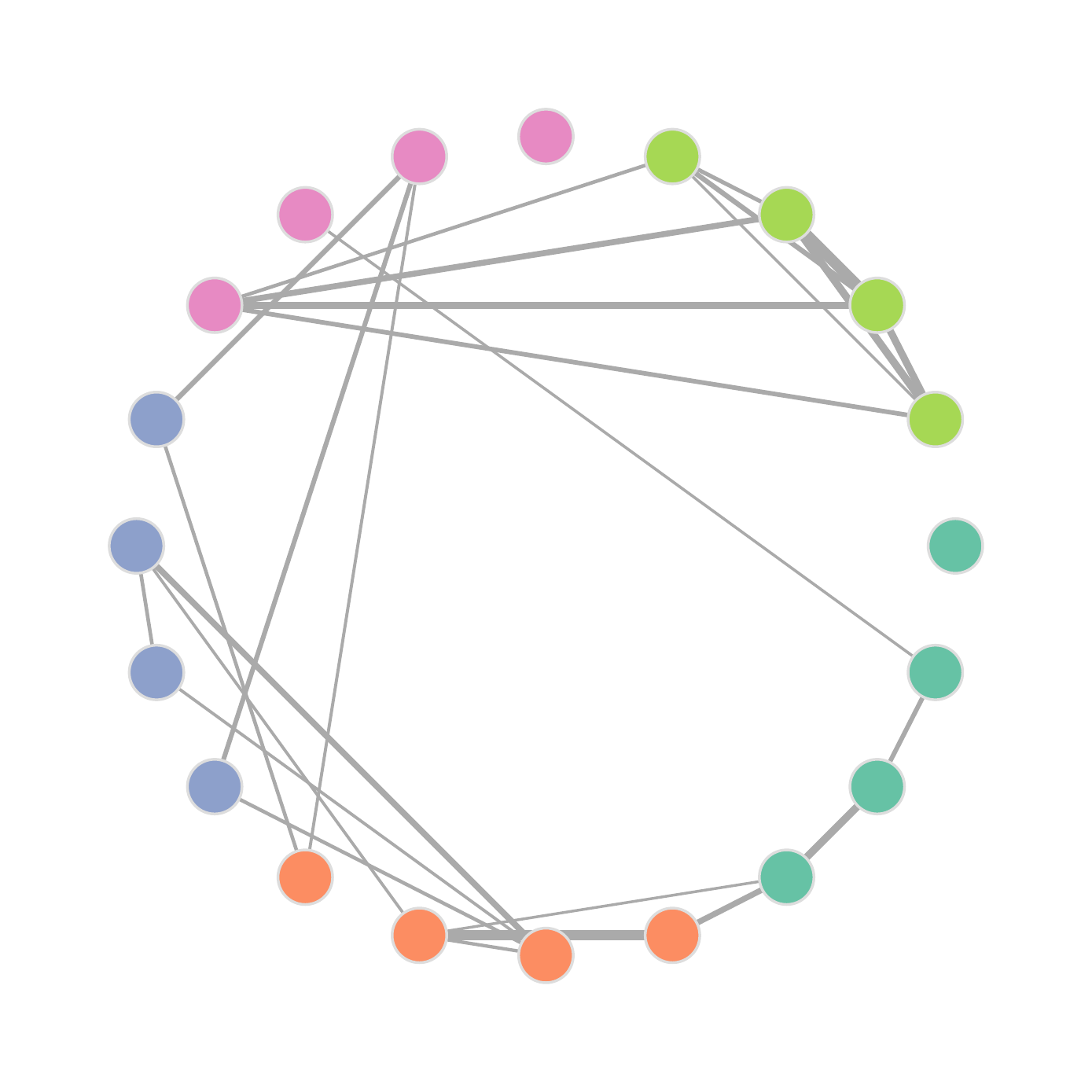}
    \end{subfigure}
    \begin{subfigure}[b]{0.2\textwidth}
        \includegraphics[width=\textwidth]{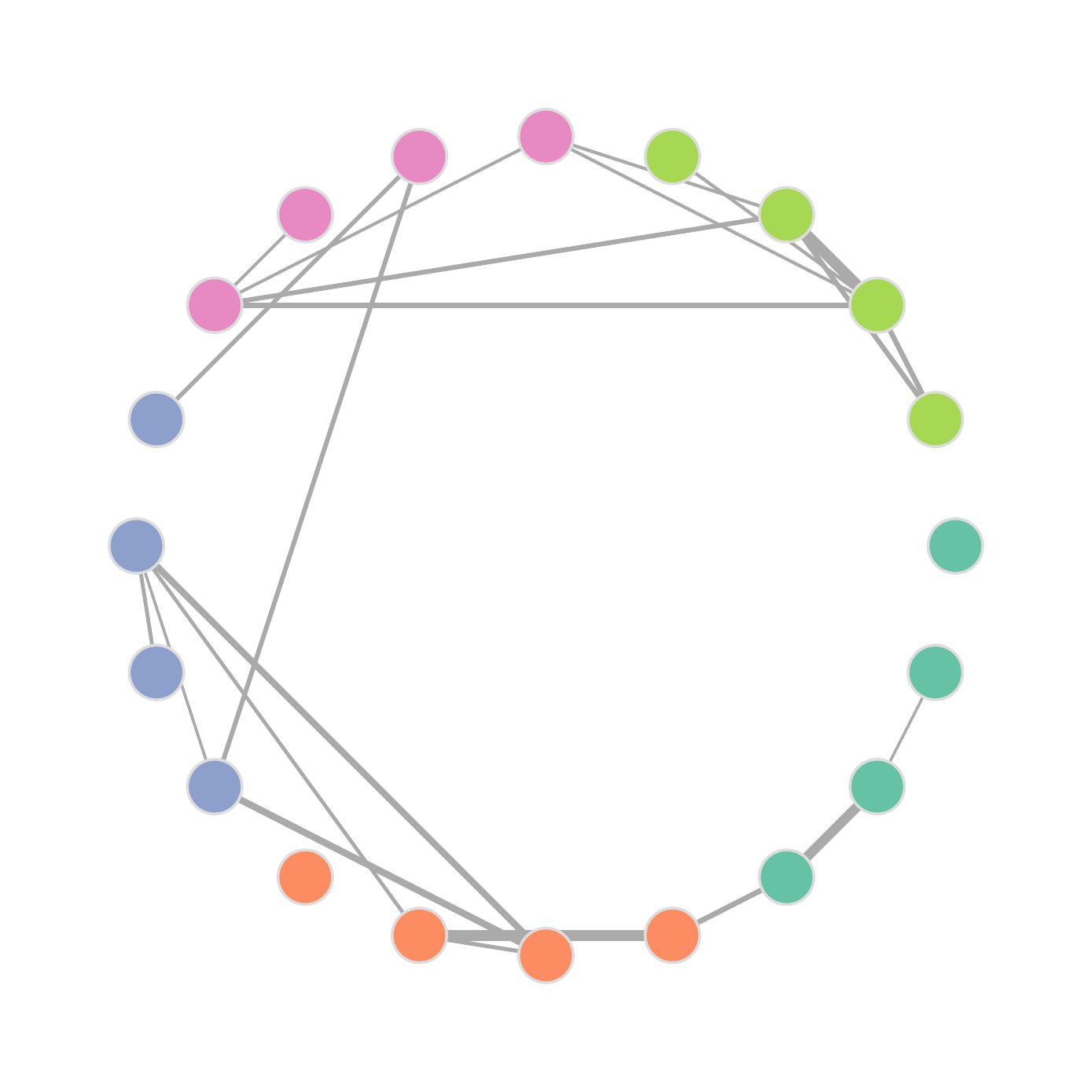}
    \end{subfigure}\\
    \begin{subfigure}[b]{0.07\textwidth}
        \includegraphics[width=\textwidth]{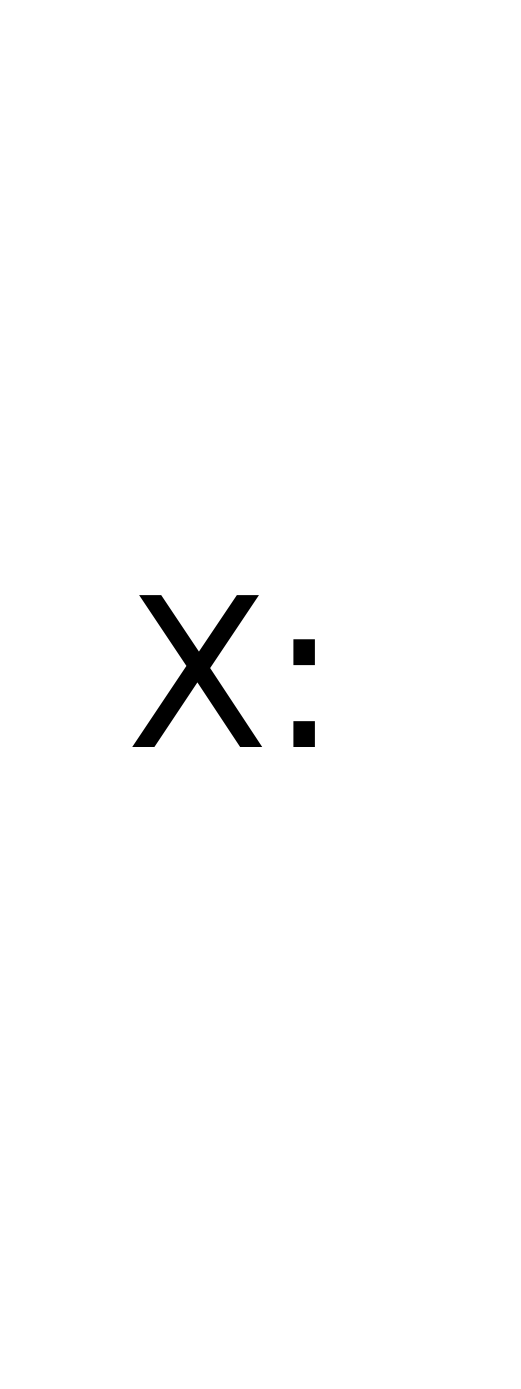}
    \end{subfigure}
    \begin{subfigure}[b]{0.2\textwidth}
        \includegraphics[width=\textwidth]{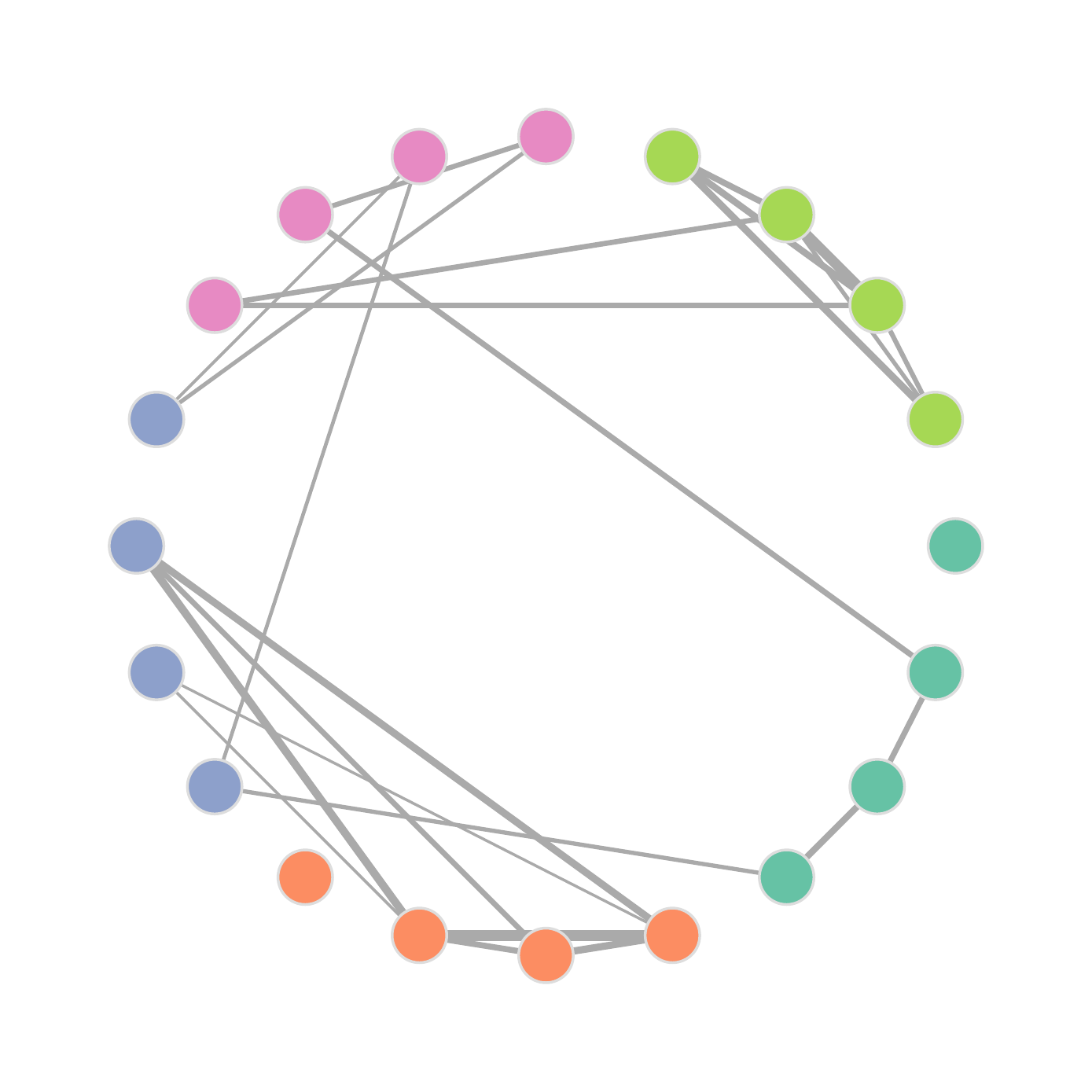}
    \end{subfigure}
    \begin{subfigure}[b]{0.2\textwidth}
        \includegraphics[width=\textwidth]{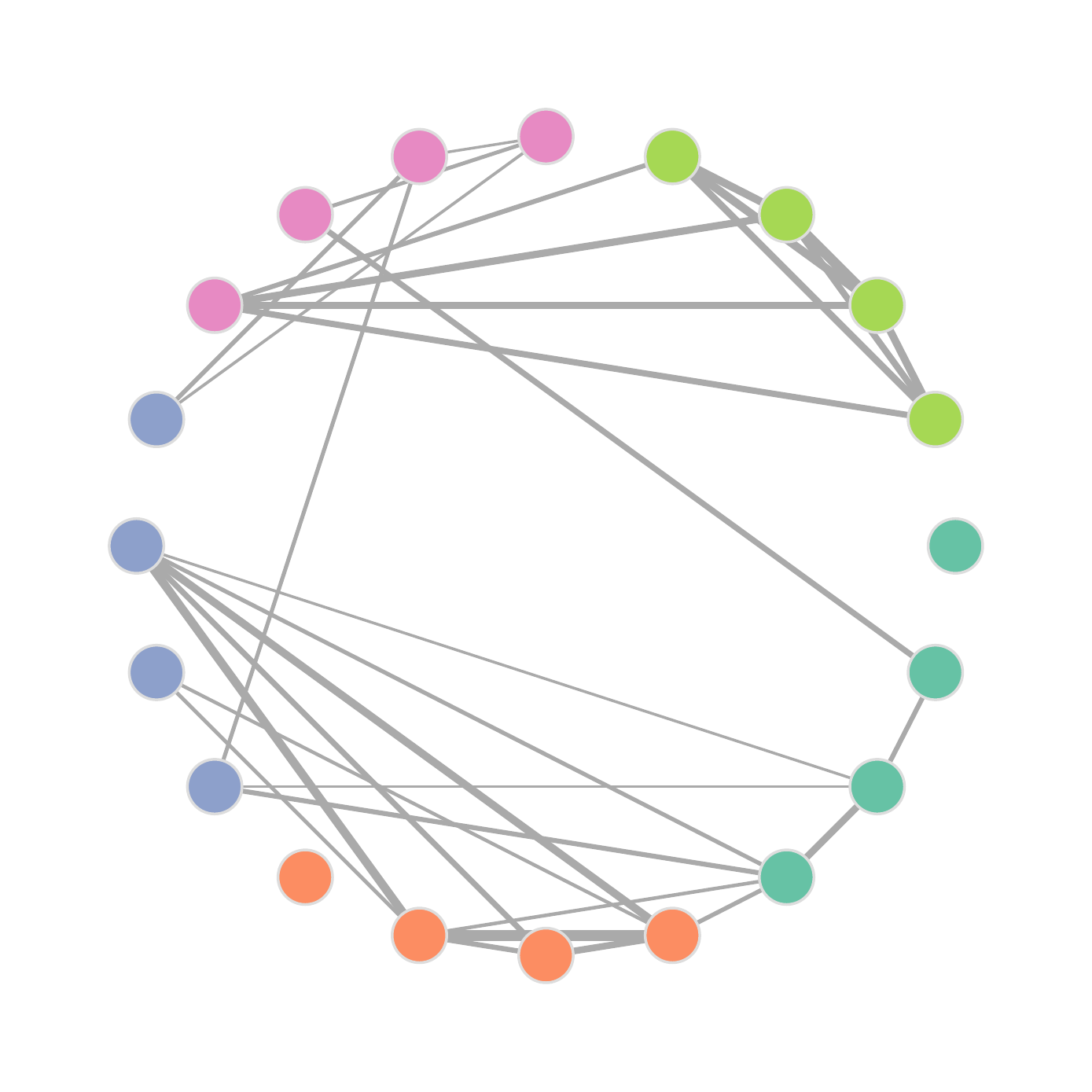}
    \end{subfigure}
    \begin{subfigure}[b]{0.2\textwidth}
        \includegraphics[width=\textwidth]{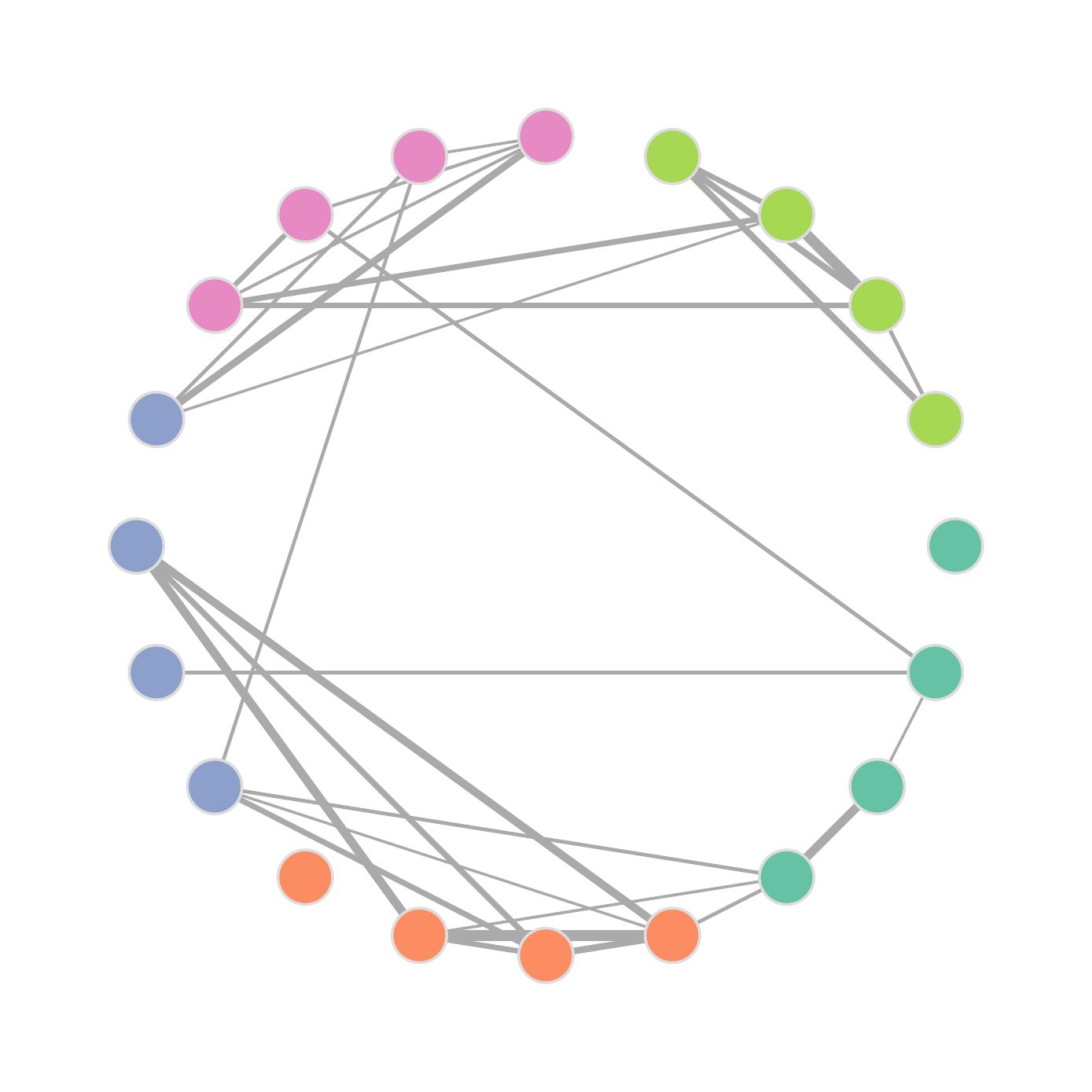}
    \end{subfigure}
    \quad\begin{subfigure}[b]{.15\textwidth}
        \centering
        \includegraphics[width=\textwidth]{figs/legend_cities.pdf}
    \end{subfigure} \\
    \begin{subfigure}[b]{0.07\textwidth}
        \includegraphics[width=\textwidth]{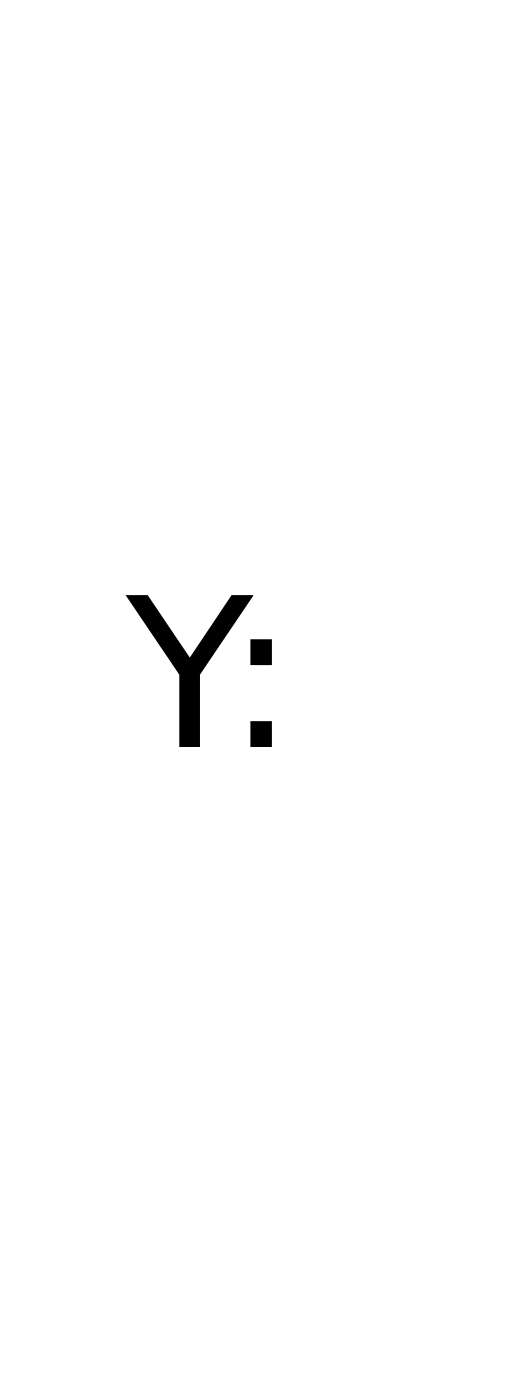}
    \end{subfigure}
    \begin{subfigure}[b]{0.2\textwidth}
        \includegraphics[width=\textwidth]{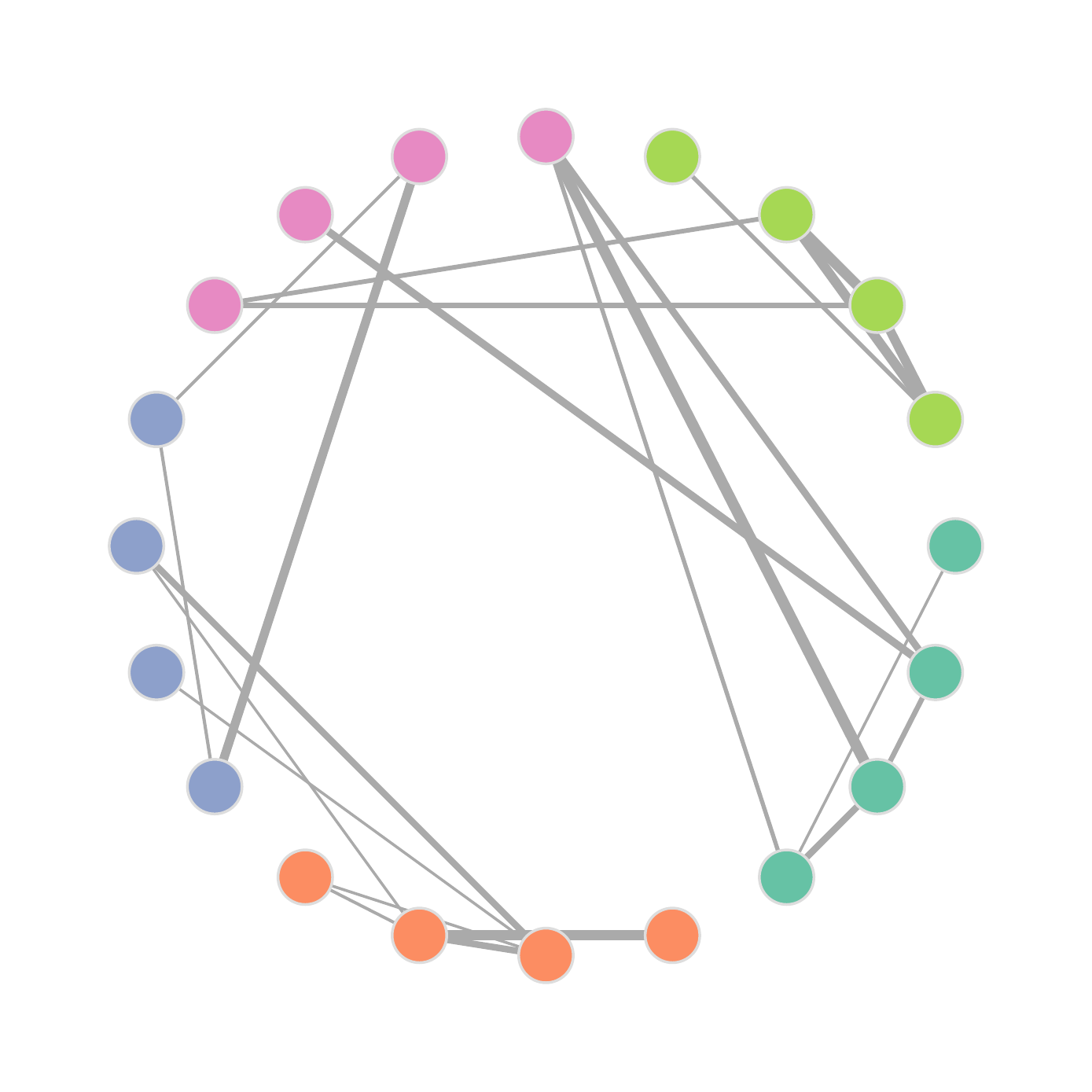}
    \end{subfigure}
    \begin{subfigure}[b]{0.2\textwidth}
        \includegraphics[width=\textwidth]{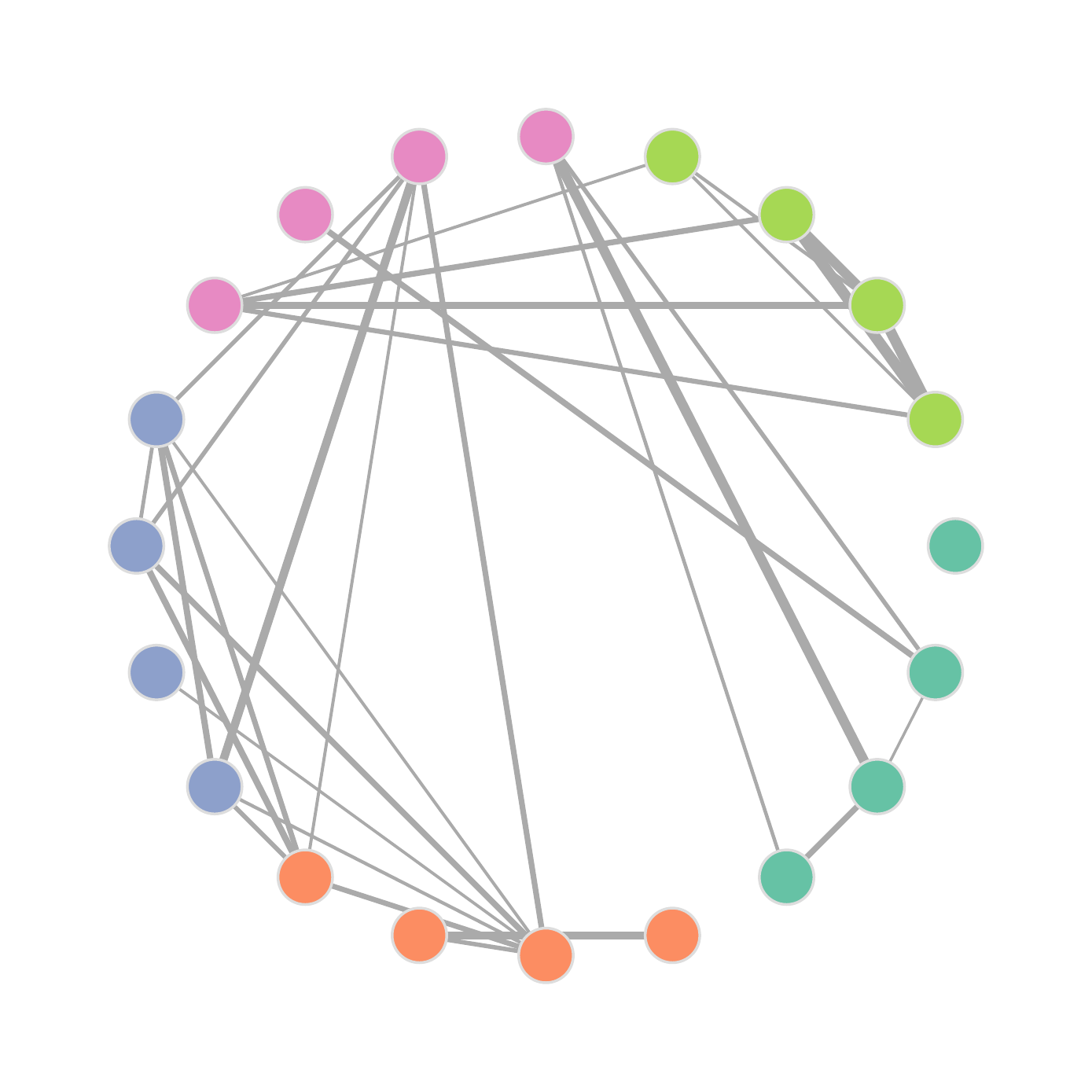}
    \end{subfigure}
    \begin{subfigure}[b]{0.2\textwidth}
        \includegraphics[width=\textwidth]{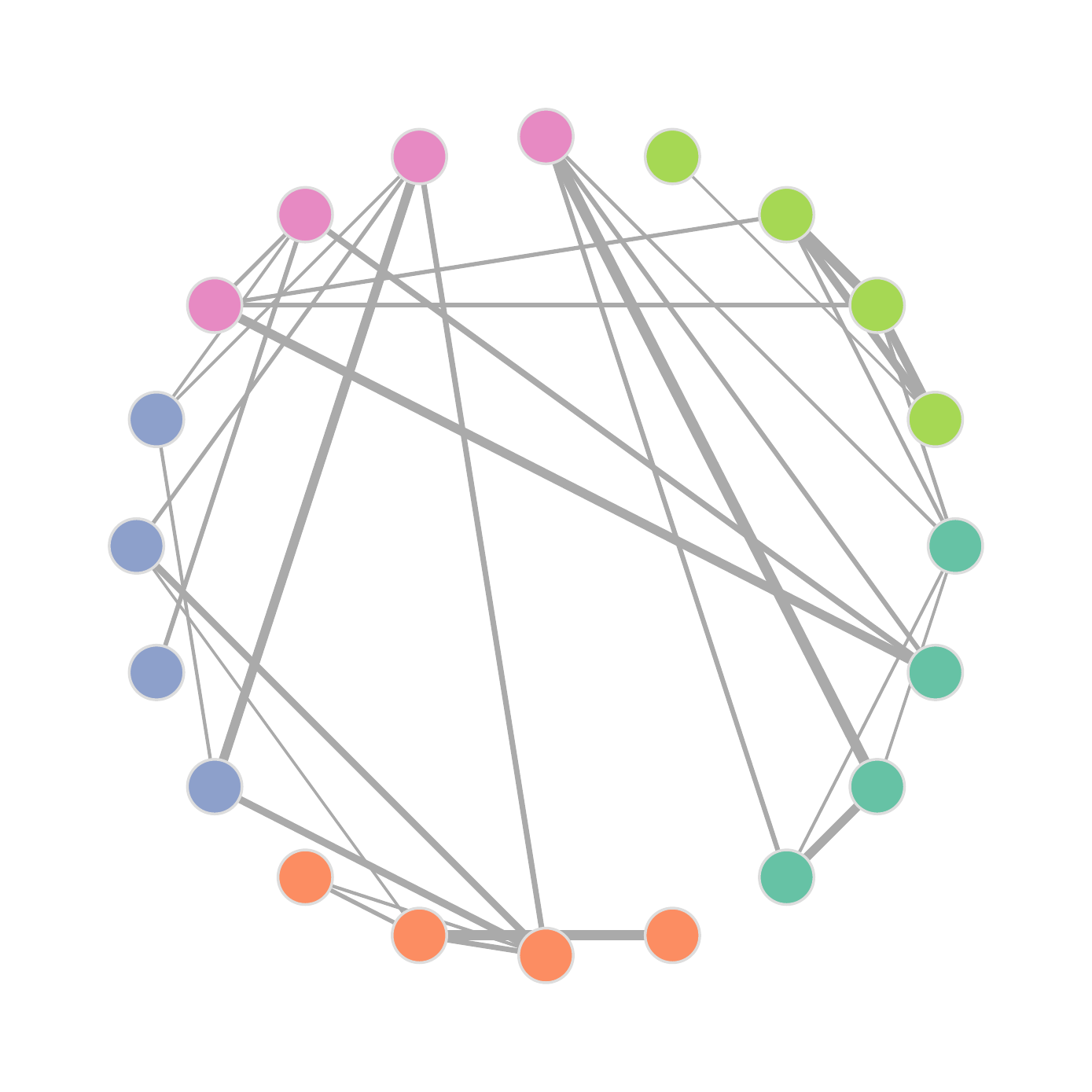}
    \end{subfigure}\\
    \begin{subfigure}[b]{0.07\textwidth}
        \includegraphics[width=\textwidth]{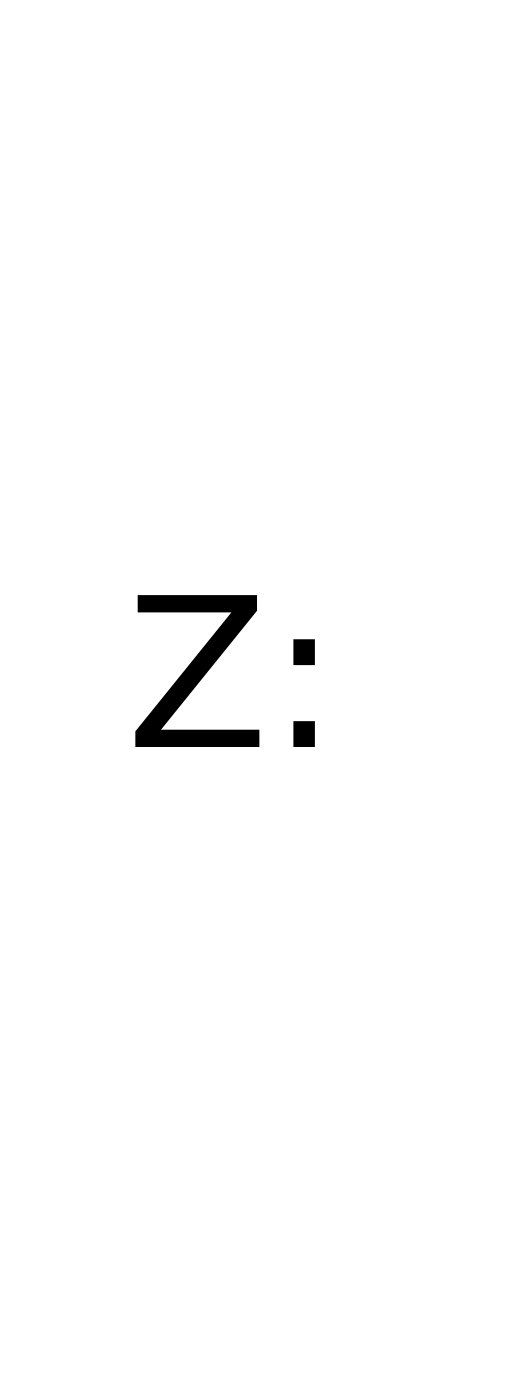}
    \end{subfigure}
    \begin{subfigure}[b]{0.2\textwidth}
        \includegraphics[width=\textwidth]{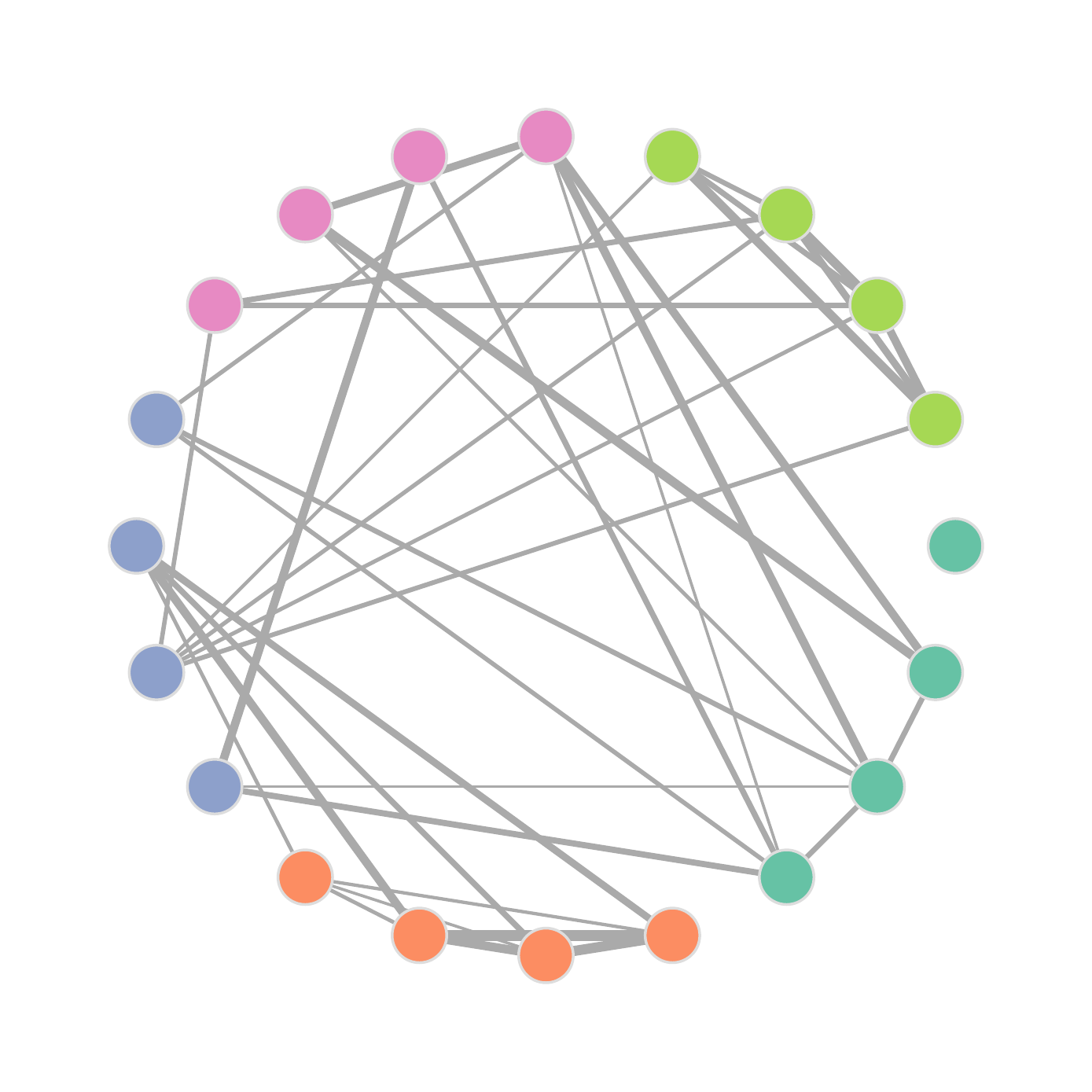}
            \centering
    \end{subfigure}
    \begin{subfigure}[b]{0.2\textwidth}
        \includegraphics[width=\textwidth]{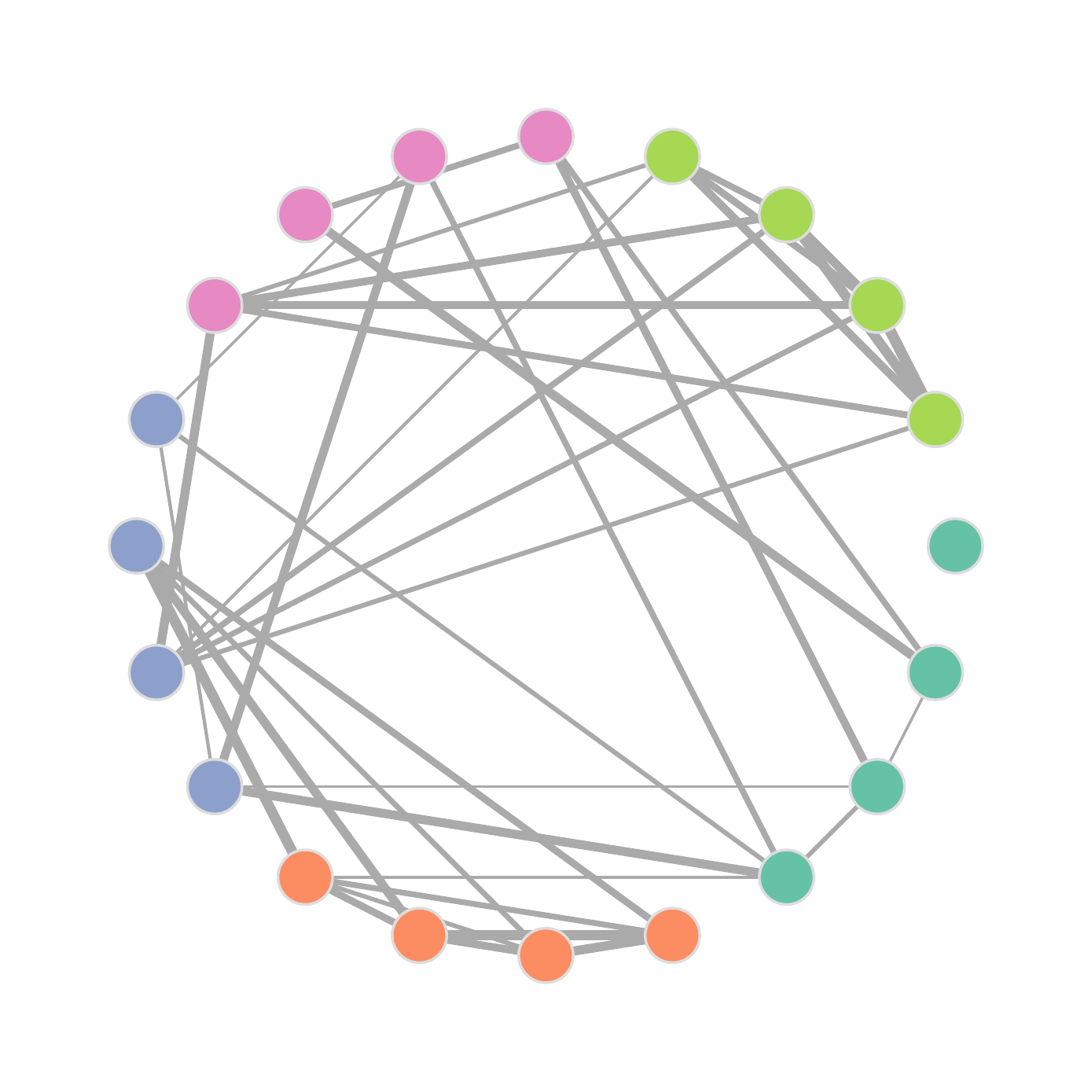}
    \end{subfigure}
    \begin{subfigure}[b]{0.2\textwidth}
        \includegraphics[width=\textwidth]{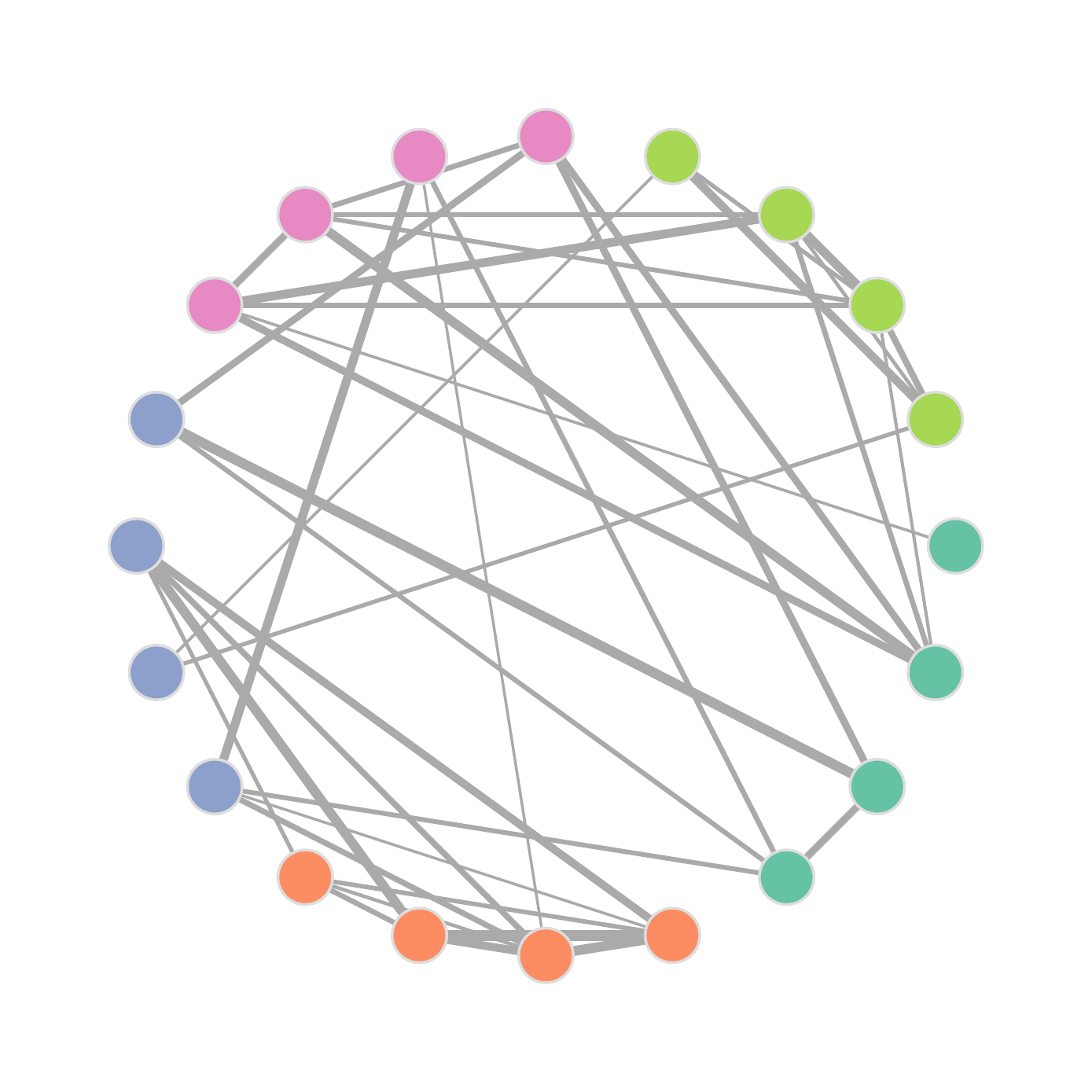}
    \end{subfigure}
    \caption{Cities networks corresponding (from left to right) to the hub, as well as 
    the nodes with the second and third highest strengths in each of the communities 
    identified in the features network. Each of the four rows corresponds to the respective
    communities I-IV in Fig.~\ref{fig:coincgraph}.  The cities networks along each of the rows present a marked mutual
    similarity, confirming further the effectiveness of the coincidence methodology in obtaining
    consistent features networks.}
        \label{fig:circles}
\end{figure}

Each of the rows in Fig.~\ref{fig:circles} presents the cities networks corresponding to the respective
prototype (hub), as well as the nodes with the second and third highest strengths
within each of the identified four communities.  For simplicity's sake, the
networks have been shown in circular format, with the nodes following the same
order as in Table~\ref{tab:cities}  Therefore, most of the adjacent nodes tend to be from the
same respective country.   The four obtained hubs correspond to the features combinations
$(1,2,3)$, $(1,2,3,4)$, $(1,2,3,5)$, and $(1,2,3,4,5)$, corresponding in the case of
the specific example to the 
nodes with maximum number of features within each respectively obtained community.
In addition, as it can be readily appreciated from Fig.~\ref{fig:circles},
the cities networks obtained for each of the four modules (rows) are remarkably similar
one another, confirming the effectiveness of the coincidence methodology for
representing the similarity between the structures of the considered networks.

\begin{figure}[ht]
\centering
    \begin{subfigure}[b]{0.24\textwidth}
        \includegraphics[width=\textwidth]{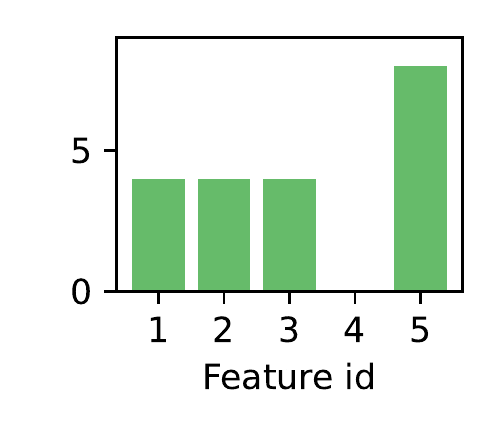}
        \caption{}
    \end{subfigure}
    \begin{subfigure}[b]{0.24\textwidth}
        \includegraphics[width=\textwidth]{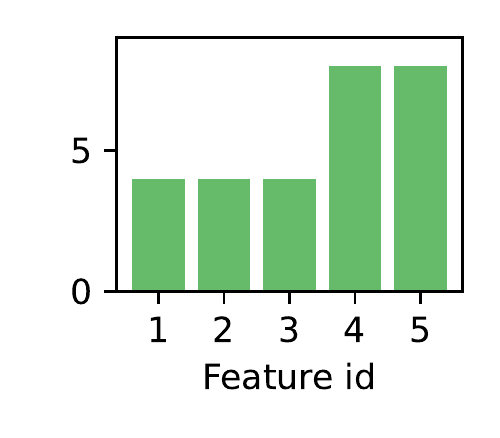}
        \caption{}
    \end{subfigure}
    \begin{subfigure}[b]{0.24\textwidth}
        \includegraphics[width=\textwidth]{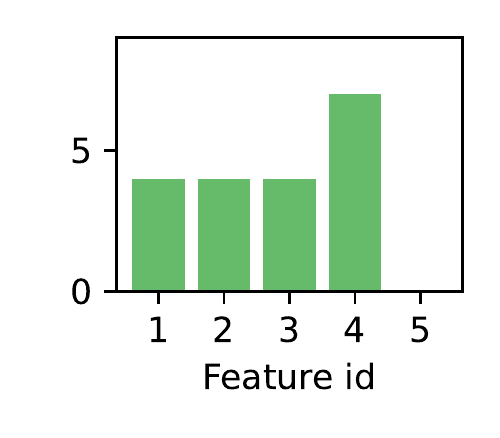}
        \caption{}
    \end{subfigure}
    \begin{subfigure}[b]{0.24\textwidth}
        \includegraphics[width=\textwidth]{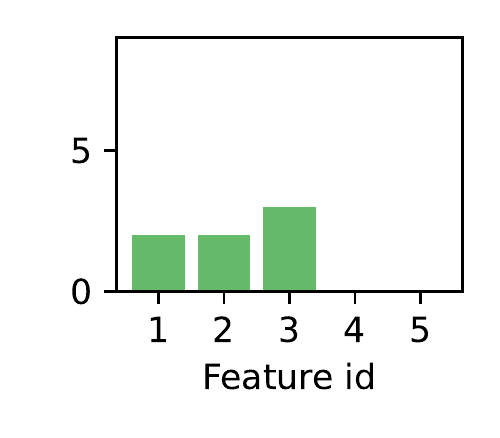}
        \caption{}
    \end{subfigure}
    \caption{Histograms of the features contained in each of the communities of the features network. From left to right, communities W (a), X (b), Y (c), and Z (d).  The three first features
    appear in all configurations, implying that the differences between the four observed models are related to the combinations of features 4 and 5.}
        \label{fig:histograms}
\end{figure}

Fig.~\ref{fig:histograms} presents histograms of how many times each feature appeared
within each of the four identified communities.  These results indicate that
the main difference between the four modules relates to features 4 and 5, namely the dispersion of point locations and standard deviation of the accessibility.
Indeed, each of the four modules are characterized by the four combinations
while taking features 4 and 5.  The histograms also indicate that  
the three other features, namely the average and standard deviation of the degree and the local transitivity, are shared by all the four main obtained modules.

\section{Concluding Remarks}

Cities can be understood as organic entities, in the sense that they are born and then keep adapting to the environment, as well as to intrinsic demands including effective transportation, basic resources and infrastructure, etc.   Given that  these two factors can vary from city to city, or even from country to country, it becomes an interesting research subject to characterize their properties while trying to establish similarity interrelationships.  In addition to contributing to a better knowledge about cities, the identification of similarities paves the way for sharing urbanistic, administrative, and planning experiences.  

The present work applied the recently introduced concept of coincidence similarity, consisting of a combination of the interiority and Jaccard index, as the means not only for transforming sets of cities characterized by respective features into respective networks but also for studying the effect of choice of these features on the obtained results.  Twenty European cities with comparable populations were selected and characterized in terms of fiver respective features, four topological and one geometric.  Except for the group of British cities, no well-defined grouping could be observed from the traditional PCA methodology.   

The coincidence methodology was then applied, and the decisive effect of its parameter $\alpha$ in obtaining detailed networks illustrated respectively to six uniformly distributed values between $0.25$ and $0.60$.
Then, by taking into account the countries of the cities as categories, we found the value of $\alpha$ that optimized the overall modularity.  Interestingly, this value (equal to $0.29$) resulted markedly distinct from the reference value of $0.5$ that would be otherwise implied in case the parameter $\alpha$ had not been taken into account.  This result corroborates the fact that modular and detailed representations of the cities could not have been obtained were not for the possibility to try different values of $\alpha$.
The maximally modular cities network obtained was characterized by four completely
separated components, each of which presenting majority of cities from a same respective country.

To complement our analysis, we applied the coincidence methodology on the weight matrices associated to the obtained cities networks while considering all possible combinations of the five adopted features.  This procedure resulted in a network whose each node corresponds to a cities network obtained by each of the possible feature combination.  Interestingly, this network resulted markedly modular, containing four well-defined communities which can be understood as the main possible data models given the adopted features. 

The obtained features network allowed interesting insights regarding the effect of the features on the cities networks.  In particular, all networks in a same of these communities will by construction have similar topology, allowing all these possible cases to be represented in terms of a model or prototype network, which was chosen to correspond to the
respective hub.  Consequently, the four identified hubs can be understood as representing the main four models representing the considered cities.  In addition, by comparing the features  in each of the four obtained communities, it has been possible to infer their respective influence on the obtained results.  In particular, we observed that all main four models share the use of the first three features, which seem to be directly related to the obtained country-specific modularity.  Interestingly, the consideration of the two remaining features then allowed a respective subdivision into the four obtained models.

In addition to their specific contributions related to cities characterization, the reported study and results also corroborated the potential of the coincidence methodology for yielding particularly detailed and modular networks when mapping datasets into networks.  It also confirmed the advantage of being able to control, through the parameter $\alpha$, the contributions of sign aligned and anti-aligned pairs of features, as a critical resource for allowing the identification of maximal modularity.  The potential of the application of the coincidence methodology for studying the effect of feature combinations was also further substantiated by the described results.

The reported results pave the way to a number of related further studies.  For instance, it would be interesting to consider other possible topological and geometrical features.  It would also be of particular interest to apply the described similarity approach to characterize parts of cities instead of their whole.  The interesting finding that European cities from a same country seem to present some uniformity could also be further investigated by considering not only additional European cities, but also samples from other continents.

\section*{Acknowledgments}
E.K.T. acknowledges FAPESP grant n. 2019/01077-3 and L. da F.C. thanks CNPq grant n. 307085/2018-0 and FAPESP 2015/22308-2 for sponsorship.  The authors also thank Filipi N. Silva, for publicly providing (GitHub) the source code for the accessibility computation.

\bibliographystyle{unsrt}
\bibliography{refs}

\end{document}